\documentclass[a4paper,11pt]{article}
\pdfoutput=1 
\usepackage{jheppub}  
\usepackage{graphicx,color}
\usepackage{amsmath}
\usepackage{mathtools}
\usepackage{amsfonts}
\usepackage{fixmath}
\usepackage{scalefnt}

\newcommand{\abbrev}{\rm\scalefont{.9}}

\newcommand{\pt}{\ensuremath{p_T}}
\newcommand{\qt}{\ensuremath{p_T}}

\newcommand{\nll}{{\abbrev NLL}}
\newcommand{\nnll}{{\abbrev NNLL}}
\newcommand{\lo}{{\abbrev LO}}
\newcommand{\nlo}{{\abbrev NLO}}
\newcommand{\nnlo}{{\abbrev NNLO}}

\newcommand{\mgen}{M}

\newcommand{\plus}{{\abbrev +}}

\newcommand{\eqn}[1]{Eq.\,(\ref{#1})}

\newcommand{\dd}{{\rm d}}

\newcommand{\higgs}{{\abbrev H}iggs}
\newcommand{\ccbar}{c\bar c}

\newcommand{\als}{\ensuremath{\alpha_s}}
\newcommand{\bob}{\ensuremath{b_0/b}}
\newcommand{\bobsq}{\ensuremath{b^2_0/b^2}}

\newcommand{\bbosq}{\ensuremath{b^2/b^2_0}}
\newcommand{\hardcoef}[2]{H_{#1}^{#2}}

\newcommand{\harda}[1]{\hardcoef{g}{A#1}}

\newcommand{\muF}{\mu_{F}}
\newcommand{\muR}{\mu_{R}}
\newcommand{\Qres}{Q}
\newcommand{\cw}{\ensuremath{\mathcal W}}

\newcommand{\ep}{\epsilon}

\newcommand{\bld}[1]{\boldmath{$#1$}}

\title{Resummed transverse momentum distribution of Pseudo-scalar Higgs boson  at NNLO$_A$+NNLL} 
\author[a]{Neelima Agarwal,}
\author[b,c]{Pulak Banerjee,}
\author[d]{Goutam Das,}
\author[b,c]{Prasanna K. Dhani,}
\author[e]{Ayan Mukhopadhyay}
\author[b,c]{V. Ravindran}
\author[e]{and Anurag Tripathi}
\affiliation[a]{Department of Physics, Chaitanya Bharathi Institute of Technology, Gandipet, Hyderabad, \\Telangana State 500075, India}
\affiliation[b]{The Institute of Mathematical Sciences, IV Cross Road, CIT Campus, Chennai 600 113, India}
\affiliation[c]{Homi Bhabha National Institute, Training School Complex, Anushakti Nagar, Mumbai 400085, India}
\affiliation[d]{Theory Group, Deutsches Elektronen-Synchrotron (DESY), Notkestrasse 85, D-22607 Hamburg, Germany}
\affiliation[e]{Department of Physics, Indian Institute of Technology Hyderabad, Kandi, Sangareddy, Telangana State 502285, India}
\emailAdd{neel1dph@gmail.com}
\emailAdd{bpulak@imsc.res.in}
\emailAdd{goutam.das@desy.de}
\emailAdd{prasannakd@imsc.res.in}
\emailAdd{ayanmukhopadhyay5@gmail.com} 
\emailAdd{ravindra@imsc.res.in}
\emailAdd{tripathi@iith.ac.in}
\abstract{In this article we have studied the transverse momentum distribution of the pseudo-scalar Higgs boson at the Large Hadron Collider (LHC). 
The small $\pt$ region which provides the bulk of the cross section is not accessible to fixed order perturbation theory due to the presence of large
logarithms in the series.
Using the universal infrared behaviour of the QCD  we resum these large logarithms up to next-to-next-to-leading logarithmic (NNLL) accuracy. We observe a significant reduction in theoretical uncertainties due to the unphysical scales at NNLL level compared to the previous order. We present the $p_T$ distribution matched to NNLO$_A$+NNLL, valid for the whole $p_T$ region and provide a detailed phenomenological study in the context of both 14 TeV and 13 TeV LHC using different choices of masses, scales and parton distribution functions  which will be useful for the search of such particle at the LHC in near future. 
}

\begin{document} 
\preprint{DESY 18-085, IITH-PH-0002/18, IMSc/2018/05/05}
\maketitle
\newcommand{\setfmfoptions}[0]{\fmfset{curly_len}{2mm}\fmfset{wiggly_len}{3mm}\fmfset{arrow_len}{3mm} \fmfset{dash_len}{2mm    }\fmfpen{thin}}

\section{Introduction}
The Standard Model (SM) of particle physics has been very successful in explaining the properties of the fundamental particles and the interactions among them. After the discovery of the Higgs boson by the ATLAS \cite{Aad:2012tfa} and the CMS \cite{Chatrchyan:2012xdj} at the Large Hadron Collider (LHC), the SM has become the most accepted theory of particle physics. The measured properties of this new boson are in full agreement with the SM predictions so far. However SM is not the complete theory of the nature as it can not describe many things including baryogenesis, neutrino masses, hierarchy problem to name a few.  Many of these issues can be addressed by going beyond the SM (BSM), often by invoking some extended sectors. A lot of the effort is recently being made towards the discovery of such new physics beyond the SM. A plethora of models exist in this context; a large class of which predicts an extended scalar sector containing multiple scalar or pseudo-scalar  Higgs particles. Extended models like the Minimal Supersymmetric SM (MSSM) or Next-to-MSSM etc., predict a larger variety of Higgs bosons which differ among each other for example by their mass, charge, CP-parity and couplings. A simple example contains an additional Higgs doublet along with the usual Higgs doublet of the SM. After the symmetry breaking this gives rise to two CP-even (scalar) Higgs bosons ($h,H$), one of which is identified with the SM Higgs boson ($h$), a CP-odd (pseudo-scalar) Higgs boson ($A$) as well as a pair of charged scalars ($H^{\pm}$). This allows phenomenologically interesting scenarios particularly with pseudo-scalar resonances. One of the important goal at the LHC Run-II is to search for such resonances which requires a precise theoretical predictions for both inclusive as well as for exclusive observables.

Similar to the SM Higgs production, the dominant production channel for $A$ is through the gluon fusion. Therefore at the LHC the large gluon flux can boost its production cross-section to a great extent. Like the Higgs boson, the leading order (LO) prediction for $A$ suffers from large theoretical uncertainties due to dependence on the renormalisation scale $\mu_R$ through the strong coupling constant. The next-to-leading order (NLO) correction \cite{Kauffman:1993nv,Djouadi:1993ji,Spira:1995rr,Spira:1993bb} is known to increase the cross section as high as $67\%$ compared to the Born contribution with scale uncertainty varying about $35\%$. This essentially calls for higher order corrections beyond NLO. The total inclusive cross section at next-to-next-to-leading order (NNLO) has been known for quite a long time \cite{Harlander:2002wh,Harlander:2002vv,Anastasiou:2002yz,Anastasiou:2002wq,Ravindran:2003um}. The NNLO correction increases the cross-section further by $15\%$ and reduces the scale uncertainties to $15\%$. To further reduce the scale dependences one has to go even higher order considering full next-to-next-to-next-to-leading order (N$^3$LO) corrections. The complexity in full N$^3$LO correction is even higher and only recently~\cite{Anastasiou:2015ema} has been obtained for the SM Higgs boson production  with infinite top mass limit which reduces the scale uncertainty to $3\%$. The large top mass approximations turned out to be a good approximation for the Higgs case and the predictions are found to be within $1\%$ \cite{Harlander:2009bw,Harlander:2009mq,Pak:2009dg} and one could expect similar behaviour in pseudo-scalar  production as well. 

The first attempt towards the N$^3$LO corrections is made through the calculation of the threshold enhanced soft-virtual (SV) corrections. For the Higgs production these are known for a long time up to N$^3$LO$_{SV}$ \cite{Anastasiou:2014vaa,Moch:2005ky,Laenen:2005uz,Ravindran:2005vv,Ravindran:2006cg,Idilbi:2005ni}. Associated production \cite{Kumar:2014uwa} and bottom quark annihilation  \cite{Ahmed:2014cha} are also known at the same accuracy. The soft gluons effect at threshold for pseudo-scalar has been computed in \cite{Ahmed:2015qda} at N$^3$LO$_{SV}$ level on the subsequent computation of its form factor at three loops \cite{Ahmed:2015qpa}. Fixed order (f.o.) cross section may however give unreliable results in certain phase space (PSP) region due to large logarithms arising from soft gluon emission and needs to be resummed to all orders. The soft gluon resummation for inclusive $h$ production has been known up to next-to-next-to-leading logarithmic (NNLO+NNLL) \cite{Catani:2003zt} accuracy for a long time. The full N$^3$LO result \cite{Anastasiou:2015ema} enables to perform soft gluon resummation at next-to-next-to-next-to-leading logarithmic (N$^3$LO+N$^3$LL)~\cite{Catani:2014uta,Bonvini:2014joa,Bonvini:2016frm} accuracy (see  also \cite{Ahmed:2015sna} for renormalisation group improved prediction.). For pseudo-scalar production, an approximate N$^3$LO$_A$ result has been matched with N$^3$LL threshold resummation in \cite{Ahmed:2016otz} (see \cite{Schmidt:2015cea,deFlorian:2007sr} for earlier works in this direction).

Recently there has been a renewed interest in the resummed improved prediction for exclusive observables as well. Higgs \cite{Banerjee:2017cfc} and Drell-Yan \cite{Banerjee:2018vvb} rapidity distributions are predicted at NNLO+NNLL accuracy resumming large threshold logarithms using double Mellin space formalism (see \cite{Sterman:1986aj,Catani:1989ne,Catani:1990rp} for earlier works)\footnote{ Also  see \cite{Laenen:1992ey, Sterman:2000pt,Mukherjee:2006uu,Bolzoni:2006ky,Bonvini:2010tp} for a different QCD approach and \cite{Ebert:2017uel,Becher:2006nr,Becher:2007ty} for SCET approach.}. The resummation in transverse momentum ($p_T$) distribution is also well studied in the past. The small $p_T$ region (defined by $p_T \ll M$, $M$ being typical hard scale of the theory) often spoils the f.o. predictions due to the presence of large logarithms of the type $ \ln(M^2/p_T^2)$. By resumming these large logarithms \cite{Dokshitzer:1978yd,Dokshitzer:1978hw,Parisi:1979se,Curci:1979bg,Collins:1981uk,Kodaira:1981nh,Kodaira:1982az,Collins:1984kg,Catani:1988vd,deFlorian:2000pr,Catani:2000vq,Zhu:2012ts,Li:2013mia}, the predictivity of the QCD can be recovered in the full PSP region for $p_T$ distribution. Such resummation of large logarithms can be obtained by exploiting the universal properties of the QCD in the infrared region \cite{Dokshitzer:1978hw,Parisi:1979se,Curci:1979bg,Collins:1981uk,Collins:1981va,Kodaira:1981nh,Kodaira:1982az,Altarelli:1984pt, Collins:1984kg,Catani:2000vq}.  Recently a powerful and elegant technique is provided using soft-collinear effective theory (SCET) by exploiting only the soft and collinear degrees of freedom in an effective field theory set up (see \cite{Bauer:2000ew,Bauer:2000yr,Bauer:2001ct,Bauer:2001yt,Beneke:2002ph}). These approaches have been applied to obtain the Higgs boson $p_T$ spectrum in gluon fusion up to NNLO+NNLL \cite{Bozzi:2007pn, Bozzi:2008bb,Catani:2010pd, Catani:2013vaa, Monni:2016ktx, Ebert:2016gcn,Stewart:2013faa, Grazzini:2015wpa, Ferrera:2016prr} and through bottom quark annihilation up to NNLO+NNLL \cite{Belyaev:2005bs,Harlander:2014hya}. Recently the $p_T$ distribution for the Higgs boson has been achieved to NNLO+N$^3$LL accuracy \cite{Bizon:2017rah,Chen:2018pzu,Bizon:2018foh}. Another approach to resum these large logarithms is through the parton shower (PS) simulations which has been also successful in recent times through the implementation in Monte Carlo generators like M{\sc ad}G{\sc raph}5$\_${\sc a}MC@NLO \cite{Alwall:2014hca}, POWHEG \cite{Alioli:2010xd} etc. mostly up to NLO+PS accuracy. However the accuracy of PS prediction is often not clear and has remained an active topic of research these days\footnote{For a recent study see \cite{Dasgupta:2018nvj} and references therein.}. In all these approaches, there is an effective matching scale (resummation scale or shower scale) which defines the infrared region and the hard region. Although its dependence is of higher logarithmic order, a suitable choice is needed to properly describe the full $p_T$ spectrum in a meaningful way. 

A clear understanding of the pseudo-scalar properties is also based on the precise knowledge of  differential observables like transverse momentum, rapidity etc. 
 For the pseudo-scalar production in association with a jet, the two-loop virtual amplitudes
can be found in \cite{Banerjee:2017faz}, which is important to predict the differential distribution.
The small $p_T$ region of the pseudo-scalar $p_T$ spectrum renders the f.o.~prediction unreliable due to the large logarithms in this PSP region. These logarithms have to be resummed in order to get a realistic distribution. This has been achieved at next-to-leading logarithmic (NLO+NLL) \cite{Field:2004tt} accuracy for a long time\footnote{Throughout this paper we take $g g \to A$ as the LO for $p_T$ distribution even though its contribution is $\sim\delta(p_T)$.} using universal infrared behaviour of QCD. The scale uncertainty in the peak region at NLO+NLL was found to be $25\%$ when the scale is varied simply by a factor of two. Along with the PDF uncertainty, the total theoretical uncertainties reach as large as $35\%$ near the peak. This necessitates the correction at the next order. In this article we extend this accuracy to NNLL. We obtained  different pieces necessary for $p_T$ resummation of a pseudo-scalar Higgs boson up to NNLL accuracy. The resummed contribution has to be matched with the f.o. in order to get a realistic distribution valid in the full $p_T$ spectrum. We use the ansatz prescribed in \cite{Ahmed:2016otz} to obtain the NNLO piece to a very good approximation and thereafter call it as NNLO$_{A}$.
         Finally the matched prediction is presented up to NNLO$_A$+NNLL accuracy for the pseudo-scalar $p_T$ spectrum in light of phenomenological study both at 14~TeV and 13~TeV LHC. 

The paper is organised as follows: in Sec.~\ref{sec2} we set up the theoretical framework for the resummation of large logarithms for small $p_T$ region relevant for pseudo-scalar Higgs boson production. In Sec.~\ref{results}, we will provide a detailed phenomenological study of the $p_T$ spectrum for different masses, scales and parton distribution functions (PDFs) relevant at the LHC. Finally we draw our conclusion in Sec.~\ref{conclusion}.
\section{Theoretical Framework} \label{sec2}
In this section we give the formula that carries out resummation and 
present the various coefficients that enter to it.

\subsection*{Resummation formula}
If we calculate the $p_T$ distribution of a colorless final state of mass $M$
and if $\pt$ is significantly smaller than $\mgen$, large logarithms of the form $\ln\left(\pt/\mgen\right)$ arise in the distribution $\dd\sigma/\dd\pt$ due to an
incomplete cancellation of soft and collinear contributions.  
At each successive order in the strong coupling constant, $\alpha_s (=g_s^2/4\pi)$, the highest power of the logarithm
that appears increases which renders 
the na\"ive perturbative expansion in $\als$ invalid 
 as $\pt\to 0$. However, factorisation of soft and collinear
radiation from the hard process allows us to resum these logarithms to all
orders in $\als$.  This factorisation 
occurs in the Fourier space conjugate to $\bold{p}_T$ called impact parameter space; the
variable conjugate to $\bold{p}_T$ is denoted by ${\bold b}$:
\begin{equation}
\begin{split}
f({\bf p}_T) = \frac{1}{(2\pi)^2}\int\dd^2 {\bf b}~ e^{-i{\bf b}\cdot {\bf
    p}_T} f({\bf b})\,,
\end{split}
\end{equation}
implying that the limit $\pt \rightarrow 0$ corresponds to $b\to
\infty$.  
The momentum conservation relates ${\bf p}_T$ to
  the sum of the transverse momenta ${\bf k}_T = \sum_i{\bf k}_{i,T}$ of the
  outgoing partons which is factorised in $\bf{b}$ space using 
\begin{equation}
\delta({\bf
    p}_T + {\bf k}_T) = \frac{1}{(2\pi)^{2}}\int\dd^2{\bf b}\exp[-i{\bf b}\cdot{\bf
      p}_T]\exp[-i{\bf b}\cdot{\bf k}_T].
\end{equation}
Using rotational invariance around the beam axis, the angular
integration can be performed which gives Bessel function $J_0$.
The distribution for low $\pt$ values compared to $M$ has the following 
behaviour which is obtained by resumming the large logarithms to all orders
in perturbation theory:
\begin{equation}
\begin{split}
\label{eq:res}
\frac{\dd \sigma^{F,\text{(res)}}}{\dd \qt^2} = \tau \int_0^{\infty}
\dd b\, \frac{b}{2} \,J_{0}(b \qt) \, W^F(b,\mgen,\tau)\,,
\end{split}
\end{equation}
here $\tau=\mgen^2/S$, and $S$ is the
hadronic centre-of-mass energy. 
The proper inclusion of terms $\pt\gtrsim M$ will be described in
Sec.~\ref{sec:matching}. Here and in what follows, the superscript $F$ is
attached to final state specific quantities.
It is convenient to consider the Mellin transform with respect to the
variable $\tau$: 
\begin{equation}
\begin{split}
W^F_N(b,\mgen) = \int_0^1\dd\tau \tau^{N-1} W^F(b,\mgen,\tau)\,,
\end{split}
\end{equation}
which has the following form for Higgs and pseudo-scalar Higgs production
\cite{Collins:1984kg,Catani:2000vq}\footnote{Throughout this paper the
  parameters that are not crucial for the discussion will be suppressed
  in function arguments.}

\begin{equation}
\begin{split}
 W^F_N(b,\mgen) &= \hat\sigma^{F,(0)}_{gg}\,
\exp \left \{- \int_{\bobsq}^{\mgen^2} \frac{\dd
   k^2}{k^2} \Big[ A_g(\als(k^2)) \ln \frac{\mgen^2}{k^2} + B_g(\als(k^2)) \Big]
 \right \}\\ & \times 
\sum_{i,j} 
\left[H_g^F C_1 C_2 \right]_{gg,ij}
\, f_{i,N}(\bob) \, f_{j,N}(\bob)\,,
\label{eq:wn}
\end{split}
\end{equation}
where $\hat\sigma^{F,(0)}_{gg}$ is the Born factor which is 
the parton level cross section at \lo{}. The function
$f_{i,N}(q)$ in \eqn{eq:wn} is the Mellin transform of the density
function $f_{i}(x,q)$ of parton $i$ in the proton, where $x$ is the
momentum fraction and $q$ the momentum transfer. The numerical constant
$b_0=2\exp\left(-\gamma_E\right)$, with Euler constant $\gamma_E= 0.5772...$,
is introduced for convenience. Unless indicated otherwise,
the renormalisation and the factorisation scales have been set to
$\muF=\muR=\mgen$. The symbolic factor $\left[H_g^F C_1 C_2 \right]$ in Eq.~(\ref{eq:wn}) has the 
following explicit form~\cite{Catani:2010pd}:
\begin{align}
\left[H_g^F C_1 C_2 \right]_{gg,ij} = H^{F}_{g;\mu_1\nu_1,\mu_2\nu_2}C^{\mu_1\nu_1}_{gi}C^{\mu_2\nu_2}_{gj}\,,
\end{align}
and the structure of partonic tensor, $C^{\mu\nu}_{gk}$, is given by
\begin{align}
\label{parten}
C^{\mu\nu}_{gk}(z;p_1,p_2,\bold{b};\alpha_s) = d^{\mu\nu}(p_1,p_2)C_{gk}(z;\alpha_s)+D^{\mu\nu}(p_1,p_2;\bold{b})G_{gk}(z;\alpha_s)\,,
\end{align}
where
\begin{align}
d^{\mu\nu}(p_1,p_2) = -g^{\mu\nu} + \frac{p_1^{\mu}p_2^{\nu}+p_2^{\mu}p_1^{\nu}}{p_1.p_2}\,,\nonumber\\
D^{\mu\nu}(p_1,p_2;\bold{b}) = d^{\mu\nu}(p_1,p_2) - 2\frac{b^{\mu}b^{\nu}}{\bold{b}^2}\,.
\end{align}
The vector $b^{\mu}=(0,\bold{b},0)$ is the two-dimensional impact parameter vector in the four-dimensional notation and $p_1$, $p_2$ are the momenta of colliding partons.
All the coefficients that appear in the resummation formula in \eqn{eq:wn} and Eq.~(\ref{parten}) have 
series expansions in $a_s=\alpha_s/4\pi$:
\begin{align}
\label{seriesexp}
C_{gi}(z;\als) &= \delta_{gi} \delta(1-z)  
+ \sum_{n=1}^{\infty} a_s^n C_{gi}^{(n)}(z), \nonumber\\
G_{gi}(z;\als) &= \sum_{n=1}^{\infty} a_s^n G_{gi}^{(n)}(z), \qquad
H_g^F(\als) = 1+\sum_{n=1}^{\infty} a_s^n H_g^{F,(n)}, \nonumber\\
A_g(\als) &= \sum_{n=1}^{\infty} a_s^n A_g^{(n)}, \qquad\quad\,\,\,\,
B_g(\als) = \sum_{n=1}^{\infty} a_s^n B_g^{(n)}\,.
\end{align}
The order at which these coefficients are
taken into account in \eqn{eq:wn} determines the {\it logarithmic
  accuracy} of the resummed cross section; LL means that all higher order coefficients except for
$A_g^{(1)}$ are neglected, NLL requires
$A_g^{(2)}$, $B_g^{(1)}$, $C_{gi}^{(1)}$, and $\hardcoef{g}{F,(1)}$,
etc. The coefficients required for the pseudo-scalar Higgs boson at NNLL accuracy will be given in Sec.~\ref{sec:H2}.

Since these resummation coefficients are process independent (i.e. they do not depend on specific final state), the coefficients $A_g$, $B_g$, and $C_{gi}$ that enter in the resummation formula
for the \higgs{} production with $H_g^{h} =1$ can be used for the pseudo-scalar production
as well. This choice of resummation coefficients will be termed as \higgs{} resummation scheme in this
paper 
({see
  \cite{Bozzi:2005wk} for details on resummation schemes}).  
The information of pseudo-scalar Higgs is 
contained in the {\it hard coefficient} $H_g^F$ and the Born factor
$\hat\sigma^{F,(0)}_{gg}$.
All resummation coefficients are known in the \higgs{} scheme up to the order required in this paper (see
Sec.~\ref{sec:H2}), with the exception of $H_g^{A,(1)}$ and $H_g^{A,(2)}$ whose evaluation through \nnlo{} will also be presented in Sec.~\ref{sec:H2}.

\par In the infinite top quark mass limit the effective Lagrangian \cite{Chetyrkin:1998mw} describing pseudo-scalar production is given by
\begin{equation}
 \mathcal{L}^{A}_{\text{eff}} = \Phi^{A}(x)\left[-\frac{1}{8}C_{G}O_{G} -\frac{1}{2}C_{J}O_{J}\right],
\end{equation} 
where the operators are defined as,
\begin{equation}
O_{G} = G^{\mu\nu}_{a}\tilde{G}_{a,\mu\nu}\equiv \epsilon_{\mu\nu\rho\sigma} G^{\mu\nu}_{a}G^{\rho\sigma}_{a},  \hspace*{2cm} O_{J} = \partial_{\mu}\left( \bar{\psi}\gamma^{\mu}\gamma_{5}\psi\right).
\end{equation}
The Wilson coefficients $C_G$ and $C_J$ are obtained by integrating
out the loops resulting from top quark. $G^{\mu\nu}_{a}$ and $\psi$ represent
gluonic field strength tensor and light quark fields, respectively.
In this study we will only consider contributions arising from 
the operator $O_G$ in the effective Lagrangian and will not  include the 
contributions arising from $O_J$ operator.
The Born cross section for the pseudo-scalar production at the parton level including the finite
top mass dependence is given by
\begin{align}
 \hat{\sigma}^{A,(0)}_{gg}(\mu_{R}^{2}) = \frac{\pi \sqrt{2} G_F}{16} a_s^2 {\cot}^2\beta ~\big|  \tau_A f(\tau_A)\big|^2.
\end{align}
Here $\tau_A = 4m_t^2/m_A^2$, $m_t$ is the $\overline{MS}$ top quark mass at scale $\mu_R$, 
$m_A$ is the mass of pseudoscalar particle and the function $f(\tau_{A})$ is given
by
\begin{align}
  \label{eq:FA}
  &f(\tau_{A}) = 
    \begin{cases}
      {\rm arcsin}^{2}\frac{1}{\sqrt{\tau_{A}}} & \tau_{A} \geq 1\,, \\
      -\frac{1}{4} \left( \ln
        \frac{1-\sqrt{1-\tau_{A}}}{1+\sqrt{1-\tau_{A}}} +i \pi
      \right)^{2} & \tau_{A} < 1\,.
    \end{cases}
\end{align}
In the above equation, $G_F$ is the Fermi constant and $\cot \beta$ is the ratio between vacuum expectation values of the Higgs doublets.
\subsection*{Perturbative expansion of resummation formula:}
Evolving the parton densities from $\bob$ to $\muF$ in
\eqn{eq:wn} (see Ref.~\cite{Bozzi:2005wk}), one can define the partonic
resummed cross section $\cw^F_{ij,N}$ through
\begin{align}
W^F_N(b,\mgen) =
\sum_{i,j} 
\cw^F_{ij,N} \left(   b, \mgen, \muF   \right) 
f_{i,N}(\muF) 
f_{j,N}(\muF) \,.
\label{eq:wcurlw}
\end{align}
From a perturbative point of view, $\cw^F$ can
be cast into the form
\begin{equation}
\begin{split}
\cw^F_{ij,N} &\left( b, \mgen, \muF \right)
=\hat\sigma^{F,(0)}_{gg}\Bigg\{ {\cal H}^F_{gg\leftarrow
  ij,N} ( \mgen, \Qres, \muF ) + \Sigma^F_{gg\leftarrow ij, N}( L ,
\mgen, \Qres, \muF ) \Bigg\}\,,
\label{eq:curlyW}
\end{split}
\end{equation}
where $L=\ln(Q^2\bbosq)$ denotes the logarithms that are being resummed
in $\cw^F$ and $\Qres$ is an arbitrary {\it resummation scale}.  While
$\cw^F$ is formally independent of $\Qres$, truncation of the
perturbative series will introduce a dependence on this scale which is,
however, of higher order. 
The $b$ dependence is contained entirely in the functions 
$\Sigma^F_{\ccbar\leftarrow ij}$ which are defined to vanish at
$L=0$; for the perturbative expansions up to \nnlo{} refer to
Ref.~\cite{Bozzi:2005wk}. The hard-collinear function ${\cal H}^F_{gg\leftarrow
  ij,N}$ depends on the coefficients $H^F_g$ and $C_{gi}$ of Eq.~(\ref{seriesexp}).
\subsection{Matching the cross section across the large and small $\pt$  regions}
\label{sec:matching}

The resummed result given in the previous section is valid at small values of transverse momentum where the logarithms of
$\pt$ are summed to all orders, and to emphasize that these results are accurate to a certain logarithmic accuracy such as
\nll{} or \nnll{} we attach a subscript to the resummed cross section: $\left({\dd\sigma^{F,\text{(res)}}}/{\dd\pt^2}\right)_{\text{l.a.}} $.  At high values of transverse momentum, fixed order results accurately describe the distribution which we will denote
by $\left({\dd\sigma^{F}}/{\dd\pt^2}\right)_{\text{f.o.}}$. To match the cross section across the entire $\pt$ region we will follow
the additive matching procedure  defined below:

\begin{align}
\label{eq:match}
\left(\frac{\dd\sigma^{F}}{\dd\pt^2}\right)_{\text{f.o.}+\text{l.a.}}=
\left(\frac{\dd\sigma^{F}}{\dd\pt^2}\right)_{\text{f.o.}}
+\left(\frac{\dd\sigma^{F,\text{(res)}}}{\dd\pt^2}\right)_{\text{l.a.}}
-\left(\frac{\dd\sigma^{F,\text{(res)}}}{\dd\pt^2}\right)_{\text{l.a.}}\Bigg|_{\text{f.o.}}.
\end{align}
At low $\pt$ the divergences in $\pt$ spectrum arising due to the fixed order result in the first term are subtracted by the last term, which is nothing but the expansion of the resummation formula in $a_s$ truncated to appropriate order. At large values of $\pt$ we can reduce
the effect of the last term by making the replacement
\cite{Bozzi:2005wk}
\begin{equation}
\begin{split}
L\to \tilde L \equiv \ln\left(\frac{Q^2b^2}{b_0^2}+1\right)\,.
\label{eq:Ldef}
\end{split}
\end{equation}
\subsection{Resummation coefficients and determination of \bld{\harda{,(2)}}}
\label{sec:H2}
Here we list down the $A^{(1)}_g$, $B^{(1)}_g$, $A^{(2)}_g$ \cite{Kodaira:1982az,Catani:1988vd} , $B^{(2)}_g$ \cite{deFlorian:2000pr,Davies:1984sp,deFlorian:2001zd}, $A^{(3)}_g$ \cite{Becher:2010tm}  coefficients along with $C_{gi}$ \cite{Davies:1984sp,deFlorian:2001zd,deFlorian:2000pr,Yuan:1991we} and $G_{gi}$ \cite{Catani:2010pd} coefficients that enter into the computation. Whenever, a coefficient is 
scheme dependent we have given it in the  \higgs{} scheme.
\begin{align}
A^{(1)}_g &= 4C_A\,, \nonumber\\
A^{(2)}_g &= 8C_A \left[
  \left(\frac{67}{18} -\frac{\pi^2}{6} \right)C_A -\frac{5}{9} n_f
  \right]\,,\nonumber\\ 
  A^{(3)}_g &= 64C_A\Bigg[
C_A^2  \left(\frac{11 \pi
   ^4}{720}-\frac{67 \pi ^2}{216}+\frac{245}{96}+\frac{11}{24}\zeta_3\right)+C_A  n_f \left(\frac{5 \pi
   ^2}{108}-\frac{209}{432}-\frac{7}{12} \zeta_3\right)\nonumber\\
   &+C_F n_f \left(-\frac{55}{96}+\frac{1}{2}\zeta_3\right)-\frac{1}{108} n_f^2
   + 8\beta_0\, \left(C_A \left(\frac{101}{216}-\frac{7}{16}\zeta_3\right)-\frac{7}{108} n_f\right)
\Bigg]\,,
\nonumber\\
  B^{(1)}_g &=
-\frac{2}{3}\, \left( 11C_A -2n_f \right)\,, \nonumber\\
 B_{g}^{(2)} &= 
16C_A^2 \left( \frac{23}{24} +\frac{11}{18}\pi^2 -\frac{3}{2}\zeta_3  \right) +\frac{1}{2}C_Fn_f 
-C_An_f \left( \frac{1}{12} + \frac{\pi^2}{9} \right) -\frac{11}{8} C_FC_A\,, \nonumber\\
C_{gg}^{(1)} &= \left[ (5+\pi^2)C_A -3C_F \right] \delta(1-z)\,,
\nonumber\\
C_{gq}^{(1)} &= 2C_F z \,,\nonumber\\
G_{gg}^{(1)} &= 4C_A \frac{1-z}{z} \,,\nonumber\\
G_{gq}^{(1)} &= 4C_F \frac{1-z}{z}\,,
\end{align}
where $\beta_0=(11\,C_A-2\,n_f)/3$, with the SU(N) QCD color factors $C_F=(N^2-1)/2N$, $C_A=N$ and $n_f=5$ is the number of active quark flavors.
The coefficients $A_g^{(i)}$, $B_g^{(1)}$, $C_{gq}^{(1)}$, $G_{gg}^{(1)}$ and $G_{gq}^{(1)}$ are scheme independent. The scheme dependent coefficients 
$B^{(2)}_g$ and $C_{gg}^{(1)}$ have been given above in \higgs{} scheme.
\section{The results: Hard coefficients and matched distributions} \label{results}
In this section we will first calculate the hard coefficients 
$H_g^{A,(1)}$ and $H_g^{A,(2)}$, then we will describe how we obtain the 
fixed order $\pt$ distribution that we need for the matching, and finally
obtain the distributions.
\subsection{Evaluation of hard coefficient}


\def\D{{\cal D}}
\def\DD{\overline{\cal D}}
\def\g{\overline{\cal G}}
\def\gm{\gamma}
\def\M{{\cal M}}
\def\ep{\epsilon}
\def\epm1{\frac{1}{\epsilon}}
\def\epm2{\frac{1}{\epsilon^{2}}}
\def\epm3{\frac{1}{\epsilon^{3}}}
\def\epm4{\frac{1}{\epsilon^{4}}}
\def\unM{\hat{\cal M}}
\def\ashat{\hat{a}_{s}}
\def\asmur{a_{s}^{2}(\mu_{R}^{2})}
\def\sigbar{{{\overline {\sigma}}}\left(a_{s}(\mu_{R}^{2}), L\left(\mu_{R}^{2}, m_{H}^{2}\right)\right)}
\def\sigbarn{{{{\overline \sigma}}_{n}\left(a_{s}(\mu_{R}^{2}) L\left(\mu_{R}^{2}, m_{H}^{2}\right)\right)}}
\def\unas{ \left( \frac{\hat{a}_s}{\mu_0^{\epsilon}} S_{\epsilon} \right) }
\def\rnM{{\cal M}}
\def\bt{\beta}
\def\cD{{\cal D}}
\def\cC{{\cal C}}
\def\ca{\text{\tiny C}_\text{\tiny A}}
\def\cf{\text{\tiny C}_\text{\tiny F}}
\def\ct{{\red []}}
\def\sv{\text{SV}}
\def\murOmu{\left( \frac{\mu_{R}^{2}}{\mu^{2}} \right)}
\def\bb{b{\bar{b}}}
\def\bt0{\beta_{0}}
\def\bt1{\beta_{1}}
\def\bt2{\beta_{2}}
\def\bt3{\beta_{3}}
\def\gm0{\gamma_{0}}
\def\gm1{\gamma_{1}}
\def\gm2{\gamma_{2}}
\def\gm3{\gamma_{3}}
\def\nn{\nonumber}
\def\l{\left}
\def\r{\right}
\def\F{{\cal F}}


The only coefficients that remain to be determined are the first and second order hard coefficients.
These can be extracted from the knowledge of form factors up to 2-loop for the pseudo-scalar. 
The unrenormalised form factors 
${\hat \F}^{A,(n)}_{g}$ up to 2-loop are given here
\begin{align}
  \label{eq:FF3}
  {\F}^{A}_{g} \equiv
  \sum_{n=0}^{2} \left[ {\hat a}_{s}^{n}
  \left( \frac{-q^{2}}{\mu^{2}} \right)^{n\frac{\epsilon}{2}}
  S_{\epsilon}^{n}  {\hat{\F}}^{A,(n)}_{g}\right] \, .
\end{align}
We present the unrenormalised results after factoring out Born term for the choice of the scale
$\mu_{R}^{2}=\mu_{F}^{2}=-q^{2}$\, as follows:
\begin{align}
  \label{eq:FF}
  {\hat \F}^{A,(1)}_{g} &= C_{A} \Bigg\{ - \frac{8}{\epsilon^2}
                          + 4 +
                          \zeta_2 
                          + 
                          \epsilon \Bigg( - 6 - \frac{7}{3} \zeta_3
                          \Bigg)
                          + \epsilon^2 \Bigg( 7  -
                          \frac{\zeta_2}{2} + \frac{47}{80}  \zeta_2^2
                          \Bigg) 
                           \Bigg\}\,,
                          \nonumber\\
  {\hat \F}^{A,(2)}_{g} &= C_{F} n_{f} \Bigg\{ - \frac{80}{3} +
                          6  \ln
                          \left(\frac{q^2}{m_t^2}\right)  + 8
                          \zeta_3 
                           \Bigg\} 
                          + 
                          C_{A} n_{f} \Bigg\{  -
                          \frac{8}{3 \epsilon^3}  + \frac{20}{9
                          \epsilon^2} + 
                          \frac{1}{\epsilon} \Bigg( \frac{106}{27} + 2 \zeta_2 \Bigg)
                          \nonumber\\
                        &- \frac{1591}{81} -
                          \frac{5}{3} \zeta_2     - \frac{74}{9}
                          \zeta_3 
                           \Bigg\}  
                          + 
                          C_{A}^2 \Bigg\{ 
                          \frac{32}{\epsilon^4}  + \frac{44}{3
                          \epsilon^3} + \frac{1}{\epsilon^2}\Bigg( - \frac{422}{9}  - 4
                          \zeta_2 \Bigg) 
                          +  
                          \frac{1}{\epsilon}\Bigg( \frac{890}{27} 
                        - 11 \zeta_2 
                          \nonumber\\&
                          +\frac{50}{3} \zeta_3 \Bigg)
                           + \frac{3835}{81} +  
                          \frac{115}{6} \zeta_2 - \frac{21}{5}
                          \zeta_2^2 + \frac{11}{9} \zeta_3  
                          \Bigg\}\,.
\end{align}
The
strong coupling constant
$a_{s} \equiv a_{s} \left( \mu_{R}^{2} \right)$ is renormalised at the
mass scale $\mu_{R}$ and is related to the unrenormalised one,
${\hat a}_{s} \equiv {\hat g}_{s}^{2}/16\pi^{2}$, through
\begin{align}
  \label{eq:asAasc}
  {\hat a}_{s} S_{\epsilon} = \left( \frac{\mu^{2}}{\mu_{R}^{2}}  \right)^{\epsilon/2}
  Z_{a_{s}} a_{s},
\end{align}
with
$S_{\epsilon} = {\rm exp} \left[ (\gamma_{E} - \ln 4\pi)\epsilon/2
\right]$
and the scale $\mu$ is introduced to keep the unrenormalised strong
coupling constant dimensionless in $d=4+\epsilon$ space-time
dimensions. The renormalisation constant $Z_{a_{s}}$ up to
${\cal O}(a_{s}^{3})$ is given by
\begin{align}
  \label{eq:Zas}
  Z_{a_{s}}&= 1+ a_s\left[\frac{2}{\epsilon} \beta_0\right]
             + a_s^2 \left[\frac{4}{\epsilon^2 } \beta_0^2
             + \frac{1}{\epsilon}  \beta_1 \right]
             + a_s^3 \left[\frac{8}{ \epsilon^3} \beta_0^3
             +\frac{14}{3 \epsilon^2}  \beta_0 \beta_1 +  \frac{2}{3
             \epsilon}   \beta_2 \right]\,.
\end{align}
The coefficients of the QCD $\beta$ function $\beta_{i}$ are given
by~\cite{Tarasov:1980au}
\begin{align}
  \beta_0&={11 \over 3 } C_A - {2 \over 3 } n_f \, ,
           \nonumber \\[0.5ex]
  \beta_1&={34 \over 3 } C_A^2- 2 n_f C_F -{10 \over 3} n_f C_A \, ,
           \nonumber \\[0.5ex]
  \beta_2&={2857 \over 54} C_A^3
           -{1415 \over 54} C_A^2 n_f
           +{79 \over 54} C_A n_f^2
           +{11 \over 9} C_F n_f^2
           -{205 \over 18} C_F C_A n_f
           + C_F^2 n_f\,,
\end{align}
where $n_f$ is the number of active light quark flavors.
The operator renormalisation is needed to remove the additional UV divergences and UV finite form factor is given by
\begin{align}
  \label{eq:FFRen}
  [\F^{A}_{g}]_{R} = Z^{A}_{g} \F^{A}_{g}\,,
\end{align}
where the operator renormalisation constant up to $\mathcal{O}(a_s^3)$ is given by
\begin{align}
  \label{eq:ZGG}
  Z^{A}_{g} &= 1 +  a_s \Bigg[ \frac{22}{3\epsilon}
              C_{A}  -
              \frac{4}{3\epsilon} n_{f} \Bigg]
              +
              a_s^2 \Bigg[ \frac{1}{\epsilon^2}
              \Bigg\{ \frac{484}{9} C_{A}^2 - \frac{176}{9} C_{A}
              n_{f} + \frac{16}{9} n_{f}^2 \Bigg\}
              + \frac{1}{\epsilon} \Bigg\{ \frac{34}{3} C_{A}^2
              \nonumber\\
            &-
              \frac{10}{3} C_{A} n_{f}  - 2 C_{F} n_{f} \Bigg\} \Bigg]
              +
              a_s^3 \Bigg[   \frac{1}{\epsilon^3}
              \Bigg\{ \frac{10648}{27} C_{A}^3 - \frac{1936}{9}
              C_{A}^2 n_{f}  + \frac{352}{9} C_{A} n_{f}^2  -
              \frac{64}{27} n_{f}^3 \Bigg\}
              \nonumber\\
            &+   \frac{1}{\epsilon^2}
              \Bigg\{ \frac{5236}{27} C_{A}^3 - \frac{2492}{27}
              C_{A}^2 n_{f}  - \frac{308}{9} C_{A} C_{F} n_{f}  +
              \frac{280}{27} C_{A} n_{f}^2  + \frac{56}{9} C_{F}
              n_{f}^2 \Bigg\}
              \nonumber\\
            &
              +  \frac{1}{\epsilon} \Bigg\{ \frac{2857}{81} C_{A}^3  -
              \frac{1415}{81} C_{A}^2 n_{f}  - \frac{205}{27} C_{A} C_{F} n_{f} +
              \frac{2}{3} C_{F}^2 n_{f} + \frac{79}{81} C_{A}
              n_{f}^2  + \frac{22}{27} C_{F} n_{f}^2 \Bigg\}
              \Bigg] \, .
\end{align}
We can obtain the hard coefficient function by removing  infrared singularities from renormalised form factor given in Eq.~(\ref{eq:FFRen}) by multiplying the IR subtraction operators \cite{Catani:2013tia}. This gives the hard function in what is called hard scheme. We would  however use
the $B$ and $C$ functions in the Higgs
scheme. Finally, hard coefficient functions can be calculated in the Higgs scheme by using following
relations \cite{Harlander:2014hya}:
\begin{align}
\harda{,(1)} &= \hardcoef{g,\text{hard}}{A,(1)} -
\hardcoef{g,\text{hard}}{h,(1)}\,, \nonumber \\
\harda{,(2)} &=
\hardcoef{g,\text{hard}}{A,(2)} -
\hardcoef{g,\text{hard}}{h,(2)}
+
\left(\hardcoef{g,\text{hard}}{h,(1)}\right)^2
-
\hardcoef{g,\text{hard}}{A,(1)}
\hardcoef{g,\text{hard}}{h,(1)}\,,
\end{align} 
where the subscript `hard' denotes hard scheme. The first and second order coefficients that appear in the expansion of the hard function
when calculated in the Higgs scheme are 
\begin{align}
H_g^{A,(1)} &= \frac{3}{2} C_F -\frac{1}{2}C_A\,,
\nonumber\\
H_g^{A,(2)} &= \frac{1}{12}C_F 
      + \frac{5}{96} C_A + \frac{41}{144} C_A n_f 
+ \left(-\frac{13}{8} + \frac{1}{4} \log \frac{m_A^2}{m_t^2}  \right) C_F n_f   
\nonumber \\
&+ \left(\frac{37}{24}+ \frac{11}{8}\log\frac{m_A^2}{m_t^2} \right) C_A C_F
      +\left(\frac{137}{288} -\frac{7}{8}\log\frac{m_A^2}{m_t^2} \right) C_A^2\,.
\end{align}
\subsection{Fixed order distribution at NNLO}

%

It has been long observed that the inclusive pseudo-scalar Higgs coefficient function can be obtained
from the scalar Higgs coefficient at each order of perturbation theory by a simple rescaling (see Eq.\ 13-16 of  \cite{Ahmed:2016otz}) after factoring out the born cross-section. The rescaling is exact at NLO; and at NNLO the correction terms do not contain scales explicitly and are suppressed by partonic $(1-z)^2$. 
The fact that at NLO the rescaling is exact, is already highly non-trivial and is a direct consequence of similarity of the two processes. At NNLO level the difference is only sub-dominant. 
We use the same scaling factor to obtain the approximate fixed order $p_T$ spectrum (denoted as NNLO$_A$) for pseudo-scalar since both the processes share similar kinematics. The only difference comes from the vertex corrections through virtual loop calculation which only affects the low $p_T$ spectrum and does not affect the very high $p_T$ tail. Thus we have obtained the approximate fixed order $p_T$ distribution for pseudo-scalar Higgs from scalar-Higgs spectrum by multiplying same rescaling factor as in Eq.\ 13 in Ref.\  \cite{Ahmed:2016otz}. 
We find that at NNLO level in the high $p_T$ tail, only the  rescaling coefficient from one lower order contributes to the $p_T$ spectrum.
In particular the contribution comes from $H_g^{A,(1)}$.
The fixed order distribution obtained this way has been matched to the NNLL resummed spectrum at low $p_T$  completely within HqT framework. In the next section we describe the detailed phenomenology for the matched $p_T$ spectrum.

\subsection{Matched distributions}
In this subsection we present the phenomenological aspects of the differential
distribution that we have obtained using our FORTRAN code,
which we created by modifying the publicly available code HqT~\cite{Bozzi:2003jy,Bozzi:2005wk,deFlorian:2011xf}. 
We studied the distributions for the LHC centre-of-mass energy both at 13 TeV and 14 TeV.
Our default choices for different quantities in this study are: \\\\
{\bf{For 14 TeV centre-of-mass energy}}, 
\begin{enumerate}
\item Pseudo-scalar mass $m_A=200$ GeV,
\item Resummation scale $Q=m_A$,
\item  MMHT 2014~\cite{Harland-Lang:2014zoa} parton density sets with the corresponding $\alpha_s$.
\end{enumerate}
{\bf{For 13 TeV centre-of-mass energy}},
\begin{enumerate}
\item Pseudo-scalar mass $m_A=200$ GeV,
\item Resummation scale $Q=m_A/2$,
\item  MMHT 2014~\cite{Harland-Lang:2014zoa} parton density sets with the corresponding $\alpha_s$.
\end{enumerate}

%
\begin{figure}[h!]
\begin{center}
    \begin{tabular}{cc}
     \mbox{\includegraphics[height=.23\textheight]{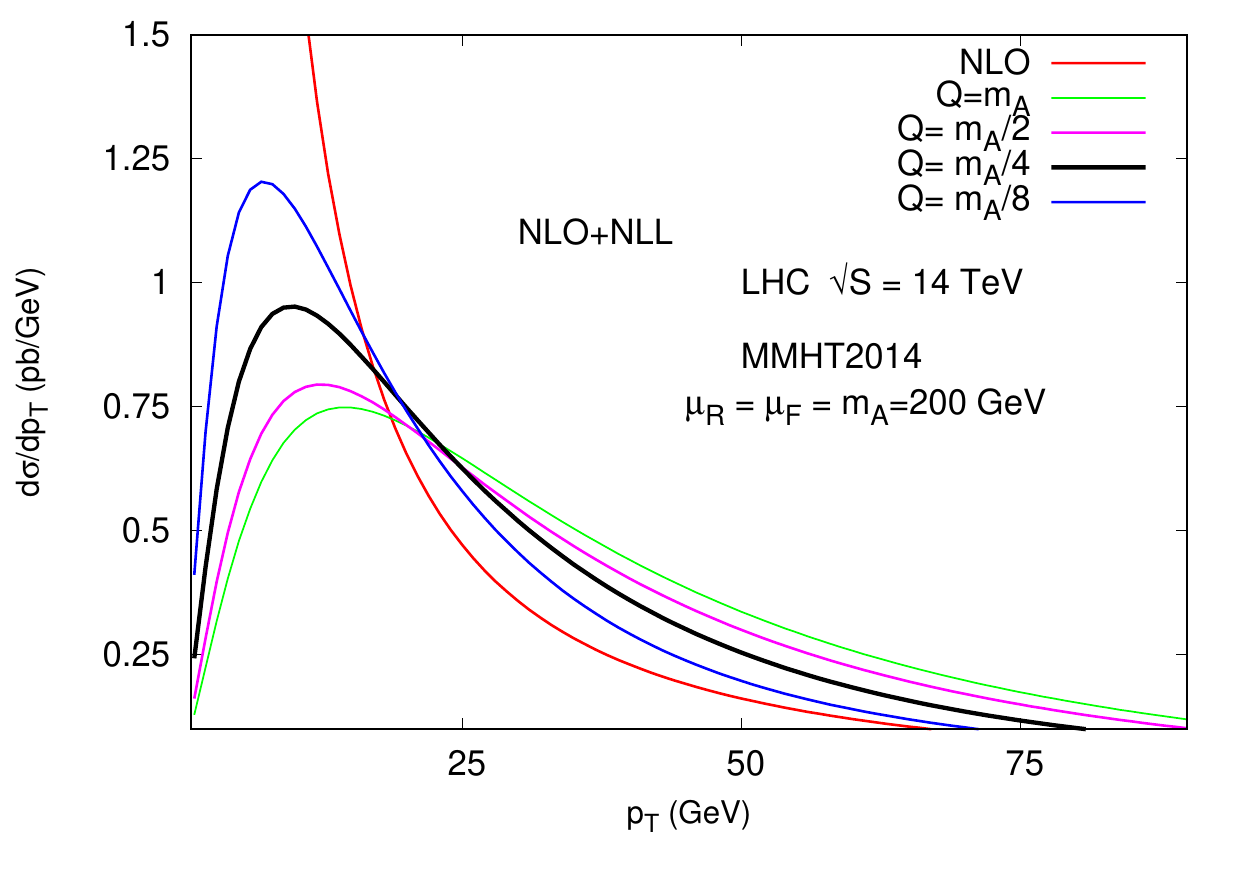}} &
     \mbox{\includegraphics[height=.23\textheight]{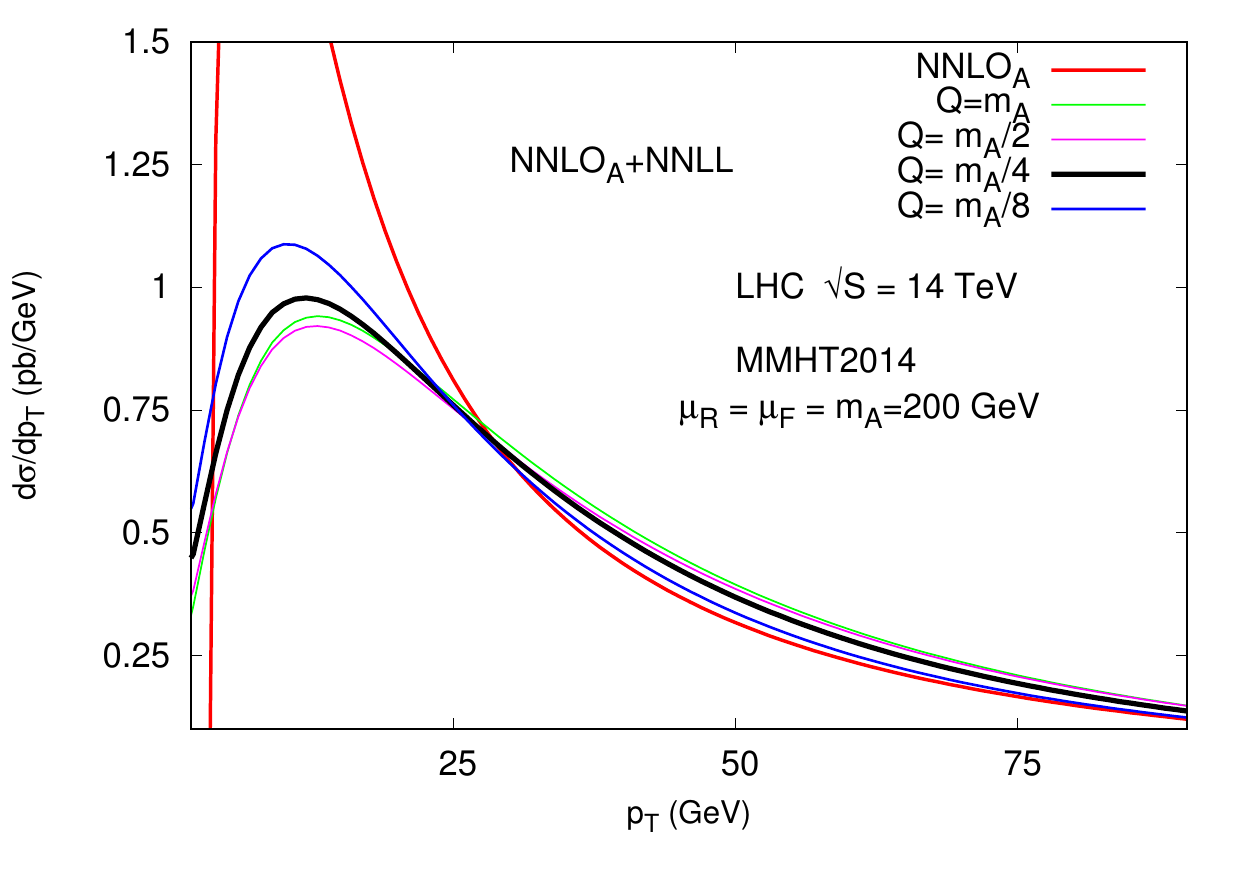}}
            \\
     \hspace{0.8cm} (a) & \qquad(b)
    \end{tabular}
    \parbox{.8\textwidth}{%
      \caption{\label{Qvar1}Resummation scale variation for (a) NLO+NLL and (b) NNLO$_A$+NNLL at 14 TeV}}
\end{center}
\end{figure}
\begin{figure}[h!]
\begin{center}
    \begin{tabular}{cc}
     \mbox{\includegraphics[height=.23\textheight]{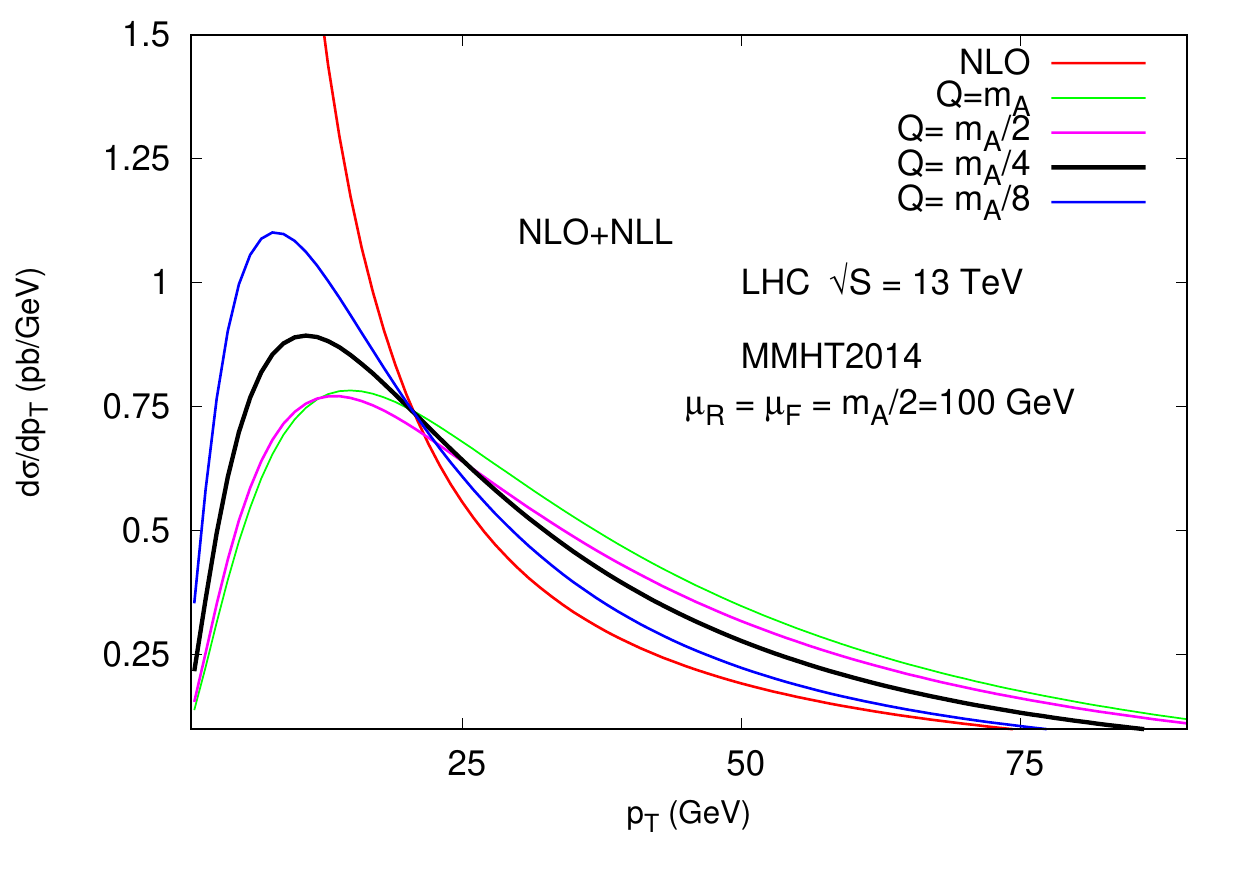}} &
     \mbox{\includegraphics[height=.23\textheight]{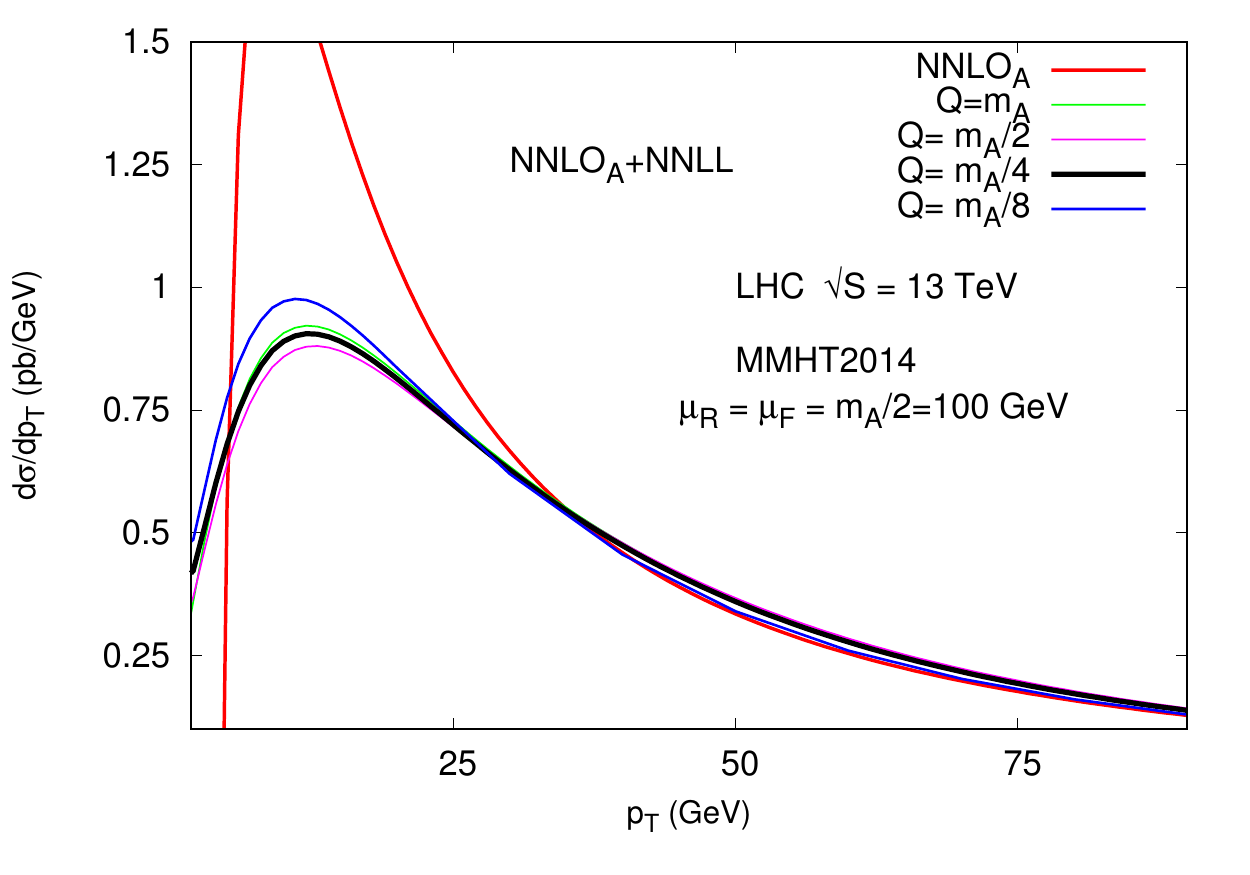}}
            \\
     \hspace{0.7cm} (a) & \qquad(b)
    \end{tabular}
    \parbox{.8\textwidth}{%
      \caption{\label{Qvar2}Resummation scale variation for
          (a) NLO+NLL and (b) NNLO$_A$+NNLL at 13 TeV}}
\end{center}
\end{figure}
In Fig.~\ref{Qvar1} (14 TeV) and Fig.~\ref{Qvar2} (13 TeV)   we study the effect of resummation over the fixed order result, where in each figure, the left 
panel shows the result for NLO and NLO+NLL; the right for NNLO$_{A}$ and NNLO$_{A}+$NNLL. For LHC 14TeV  we set 
$\mu_R=\mu_F = m_A$; for LHC 13 TeV we keep $\mu_R=\mu_F = m_A/2$  and use MMHT2014 PDF sets for both the cases. We observe that the divergent behaviour of the distribution at fixed order is cured upon resummation. Precisely, at \nlo{} the distribution diverges to positive infinity and 
at \nnlo{}$_ A$ to negative infinity. Upon resummation a regular behaviour is displayed in both the cases.

\begin{figure}[h!]
\begin{center}
    \begin{tabular}{cc}
     \mbox{\includegraphics[height=.25\textheight]{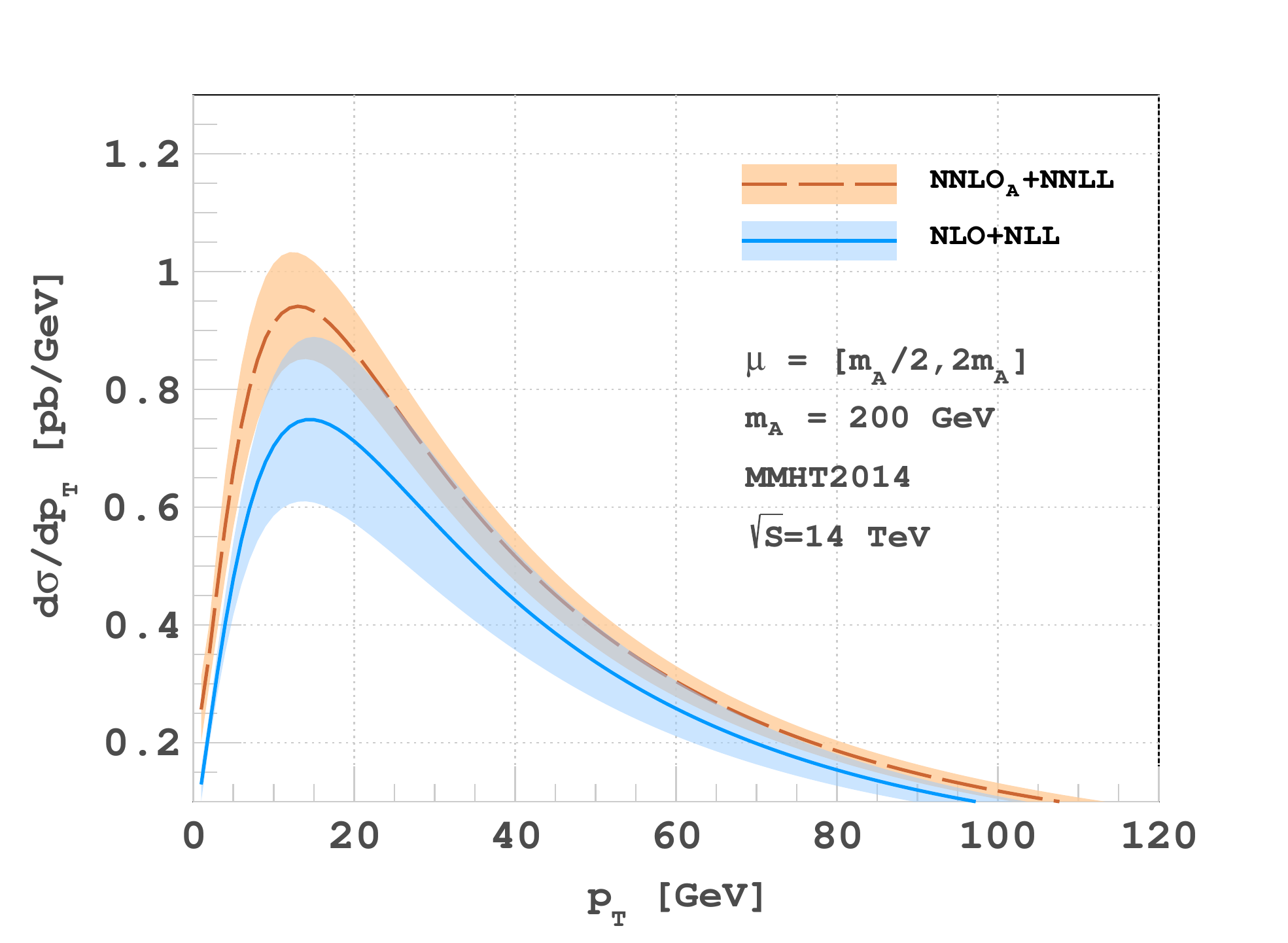}} &
     \mbox{\includegraphics[height=.25\textheight]{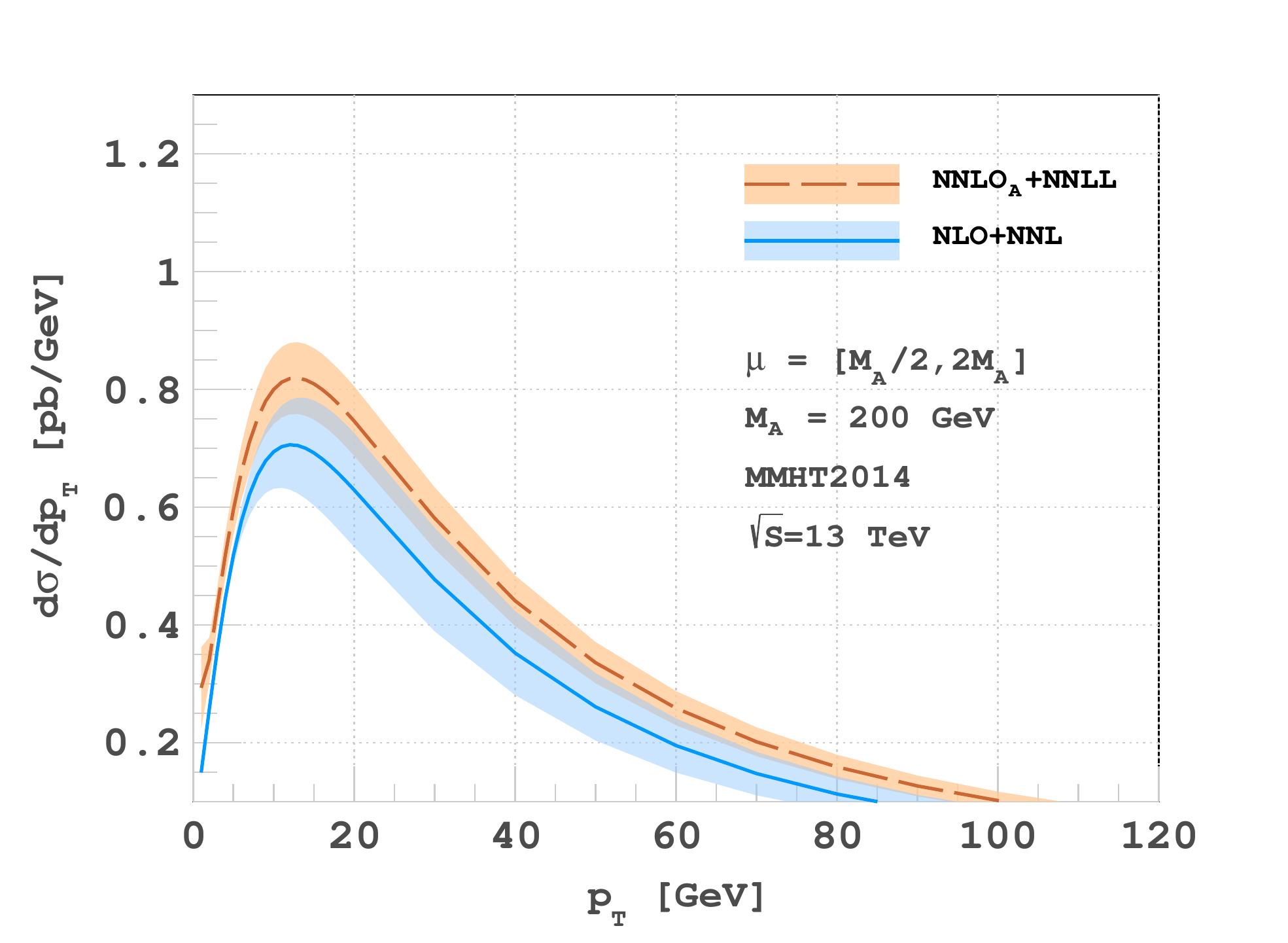}}
            \\
     \hspace{0.7cm} (a) &\qquad (b)
    \end{tabular}
    \parbox{.8\textwidth}{%
    \caption{\label{figure2}$\mu_R$ and $\mu_F$ variation at  NLO+NLL  and NNLO$_A$+NNLL for (a) 14 TeV and (b) 13 TeV}}
\end{center}
\end{figure}

\begin{figure}[h!]
\begin{center}
    \begin{tabular}{cc}
     \mbox{\includegraphics[height=.25\textheight]{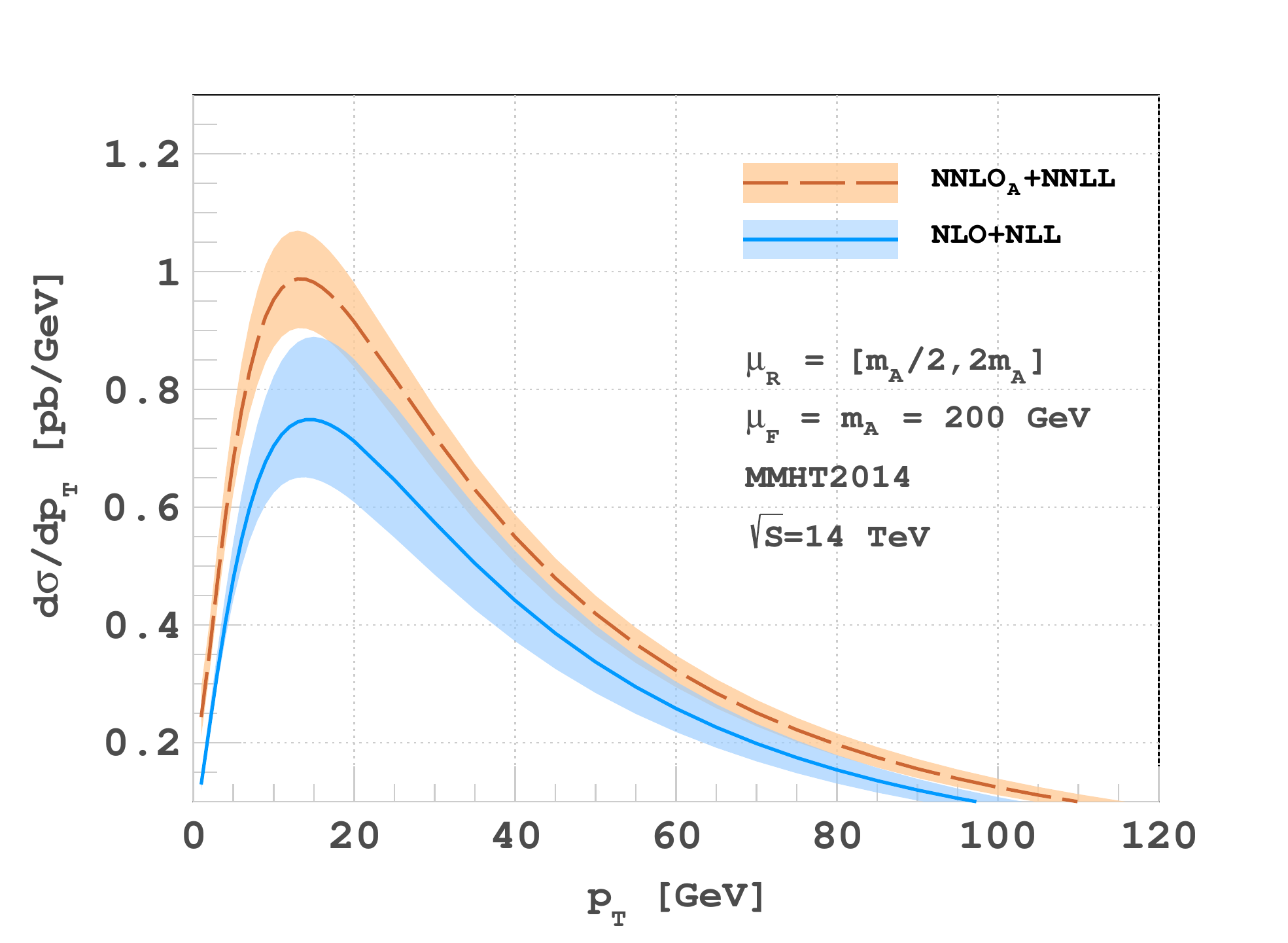}} &
     \mbox{\includegraphics[height=.25\textheight]{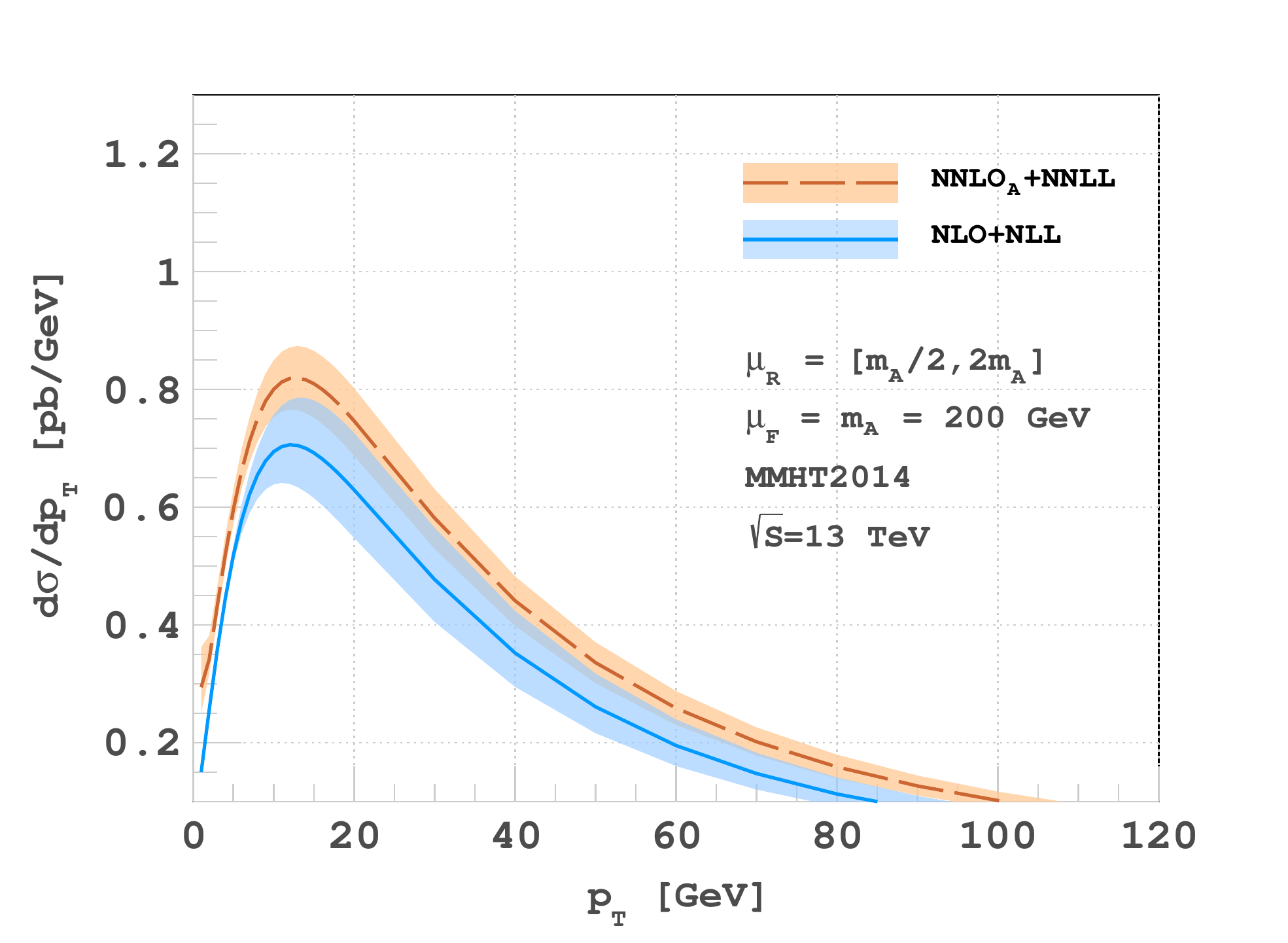}}
            \\
     \hspace{0.7cm} (a) &\qquad (b)
    \end{tabular}
    \parbox{.8\textwidth}{%
      \caption{\label{figure3}Variation of $\mu_R$ at  NLO+NLL and NNLO$_A$+NNLL keeping $\mu_F$ fixed at $m_A$ for (a) 14 TeV and (b) 13 TeV}}
\end{center}
\end{figure}

\begin{figure}[h!]
\begin{center}
    \begin{tabular}{cc}
     \mbox{\includegraphics[height=.25\textheight]{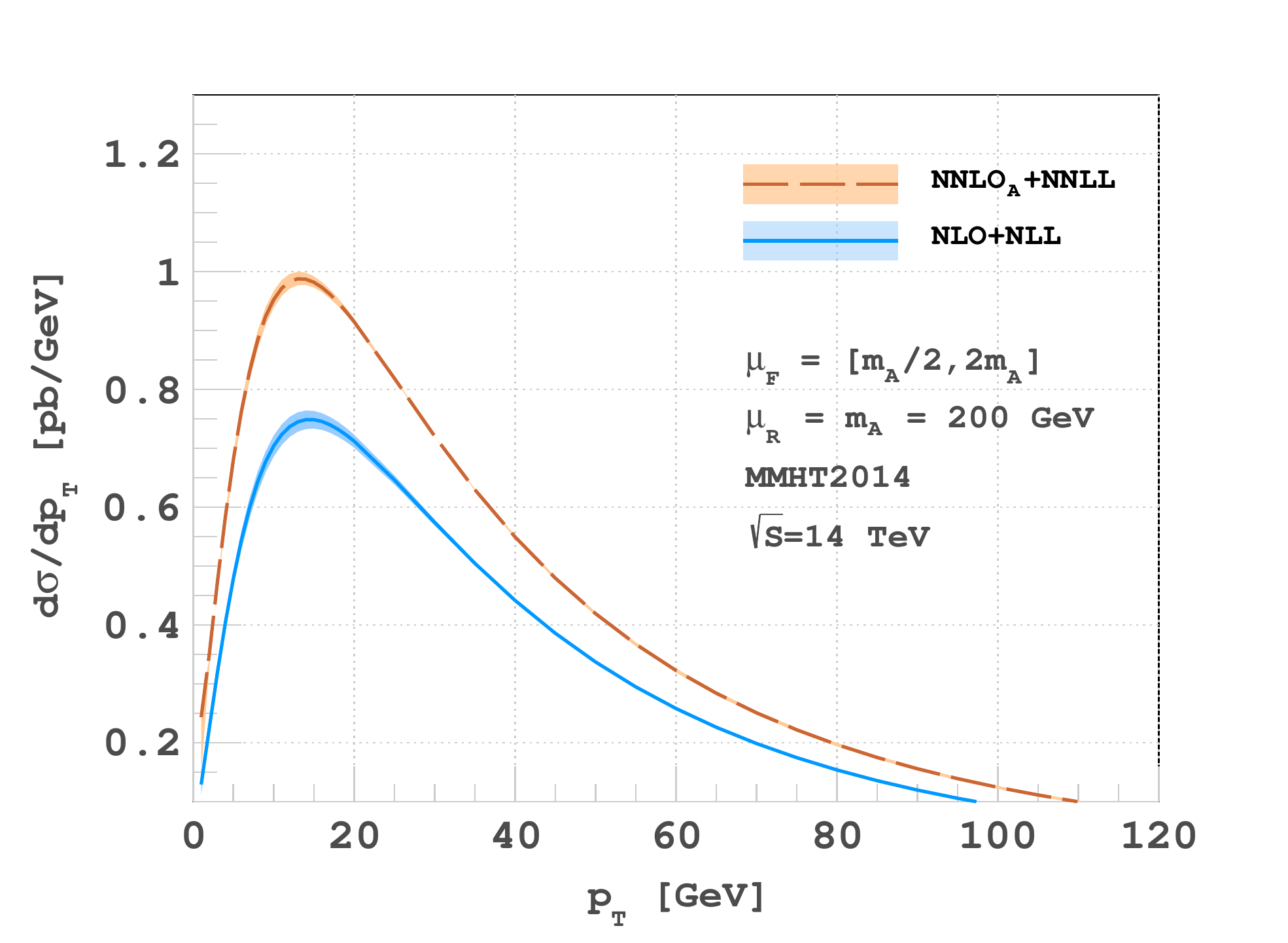}} &
     \mbox{\includegraphics[height=.25\textheight]{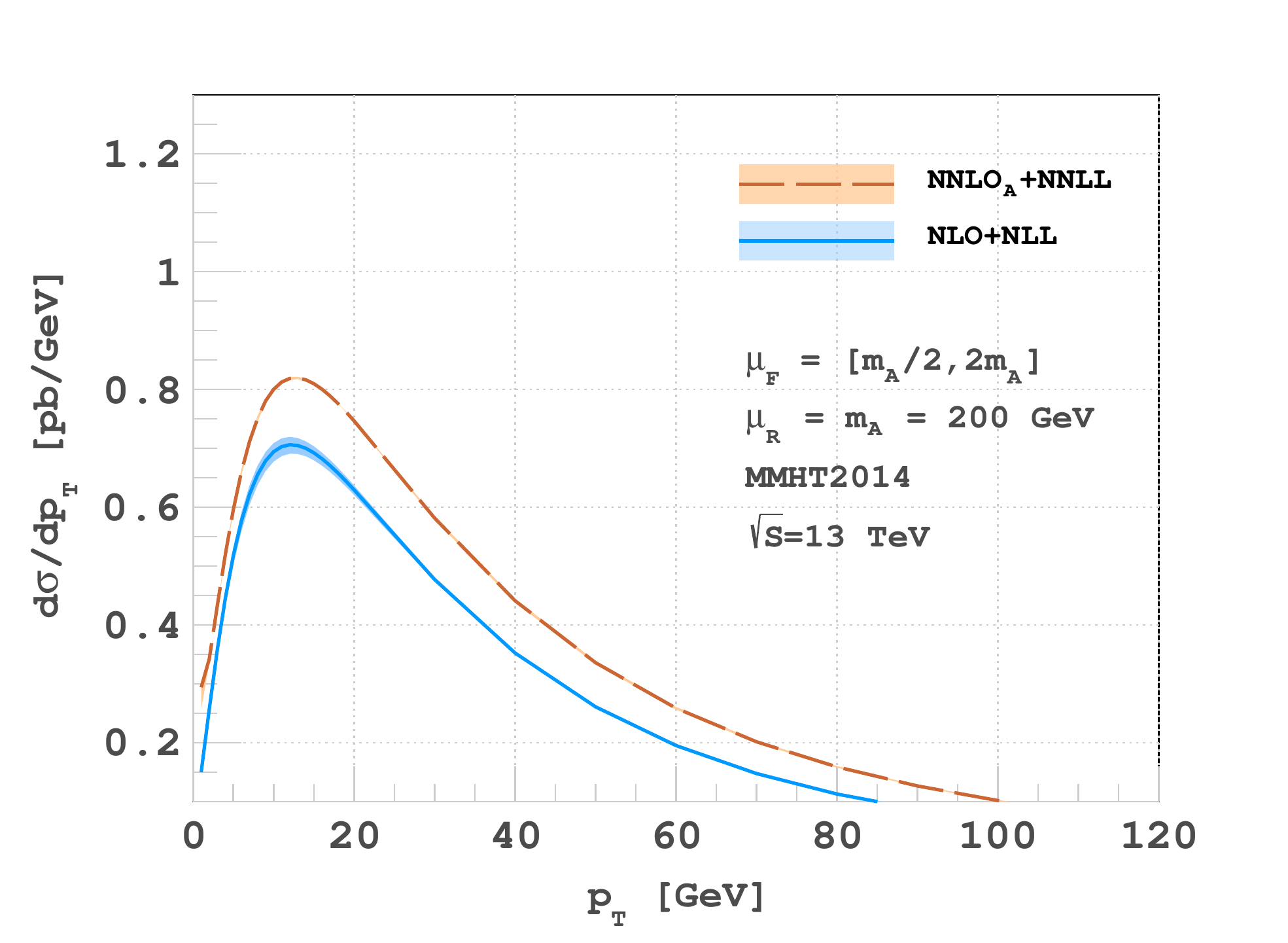}}
            \\
     \hspace{0.7cm} (a) &\qquad (b)
    \end{tabular}
    \parbox{.8\textwidth}{%
      \caption{\label{figure4}Variation of $\mu_F$ at NLO+NLL and NNLO$_A$+NNLL keeping $\mu_R$ fixed at $m_A$ for (a) 14 TeV and (b) 13 TeV}}
\end{center}
\end{figure}

\begin{figure}[h!]
\begin{center}
    \begin{tabular}{cc}
     \mbox{\includegraphics[height=.25\textheight]{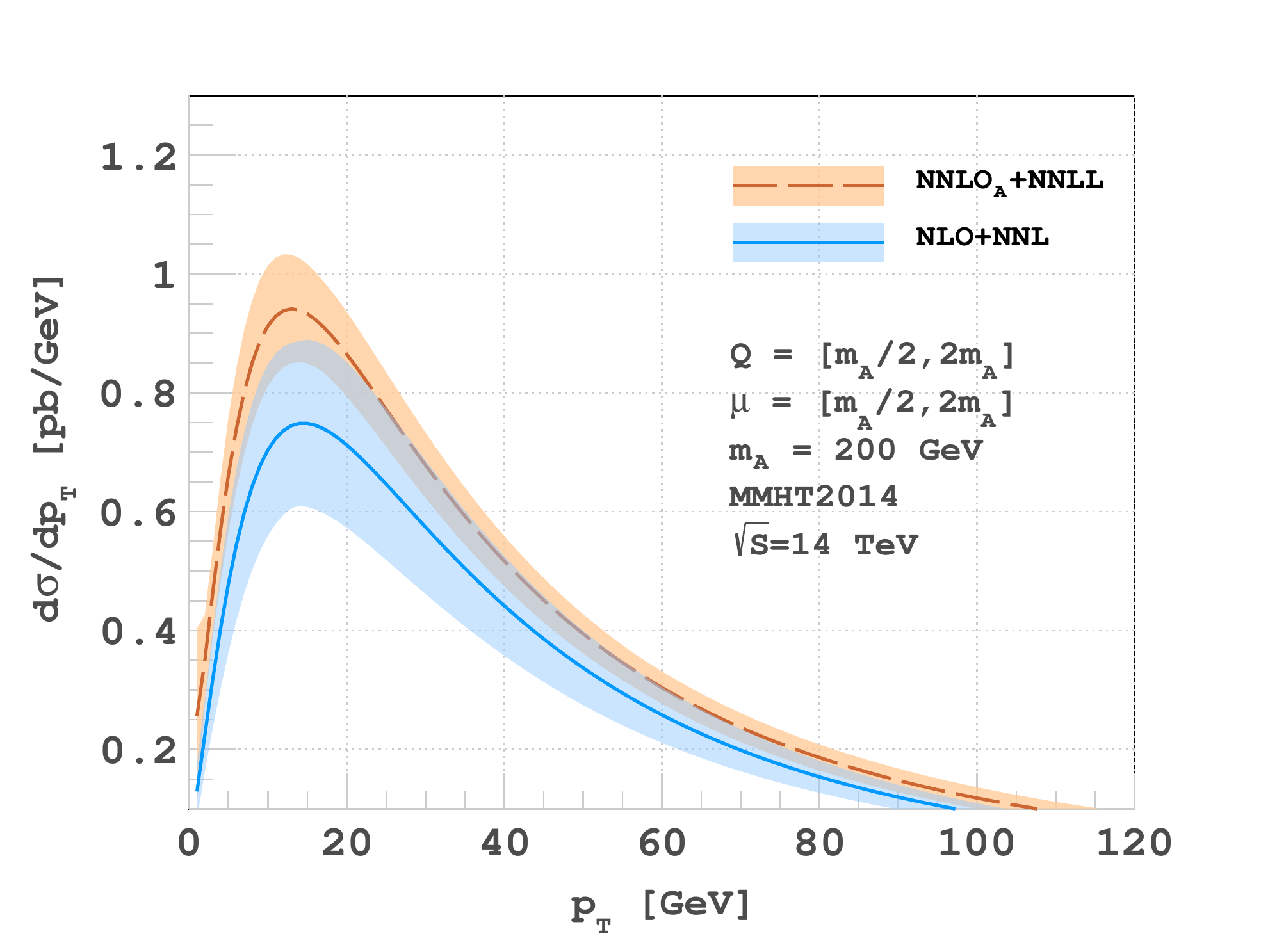}} &
     \mbox{\includegraphics[height=.25\textheight]{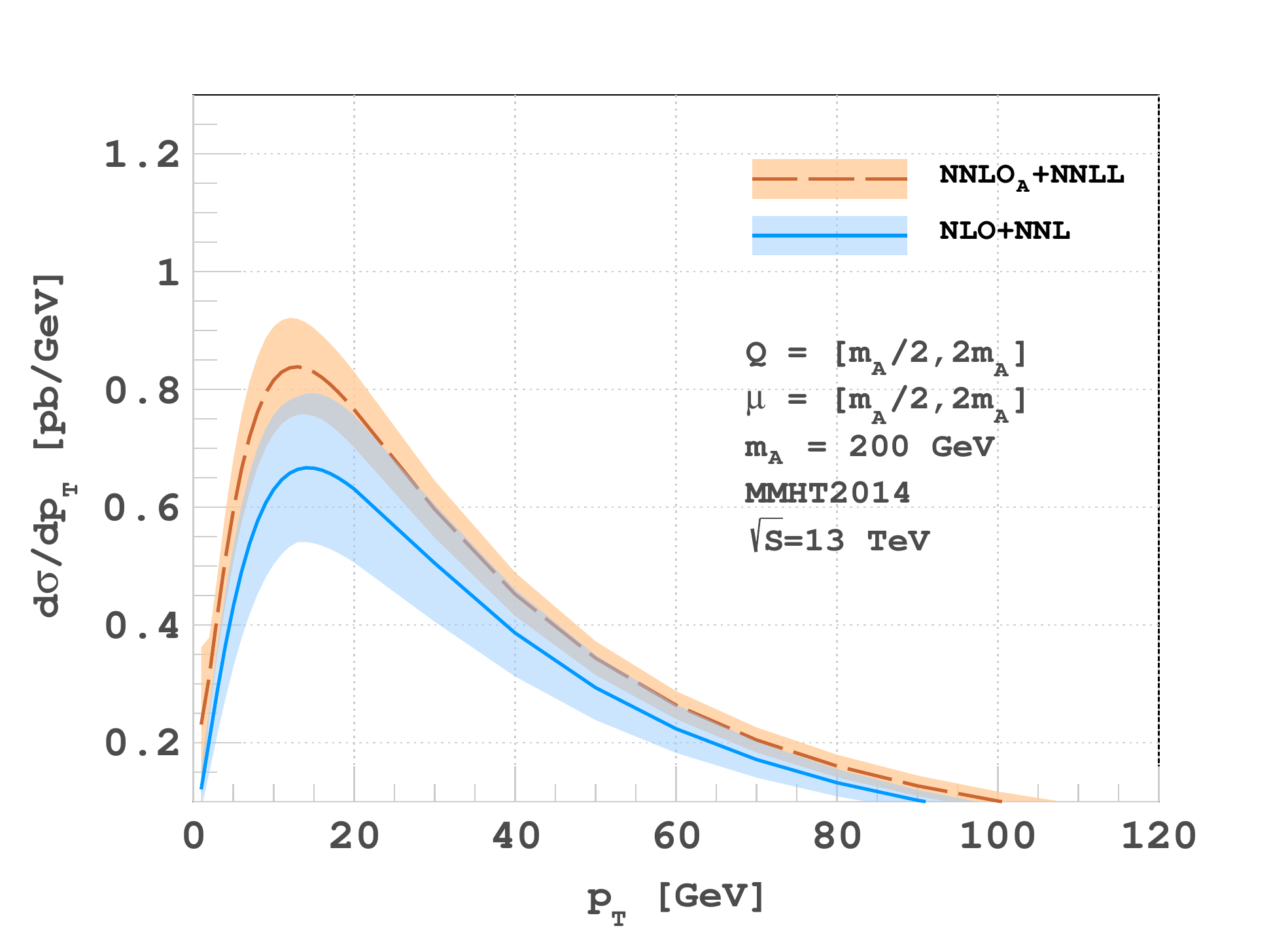}}
            \\
     \hspace{0.7cm} (a) &\qquad (b)
    \end{tabular}
    \parbox{.8\textwidth}{%
      \caption{\label{figure5}Q, $\mu_R$ and $\mu_F$ variation at NLO+NLL and NNLO$_A$+NNLL for (a) 14 TeV and (b) 13 TeV}}
\end{center}
\end{figure}

\begin{figure}[h!]
\begin{center}
    \begin{tabular}{cc}
     \mbox{\includegraphics[height=.25\textheight]{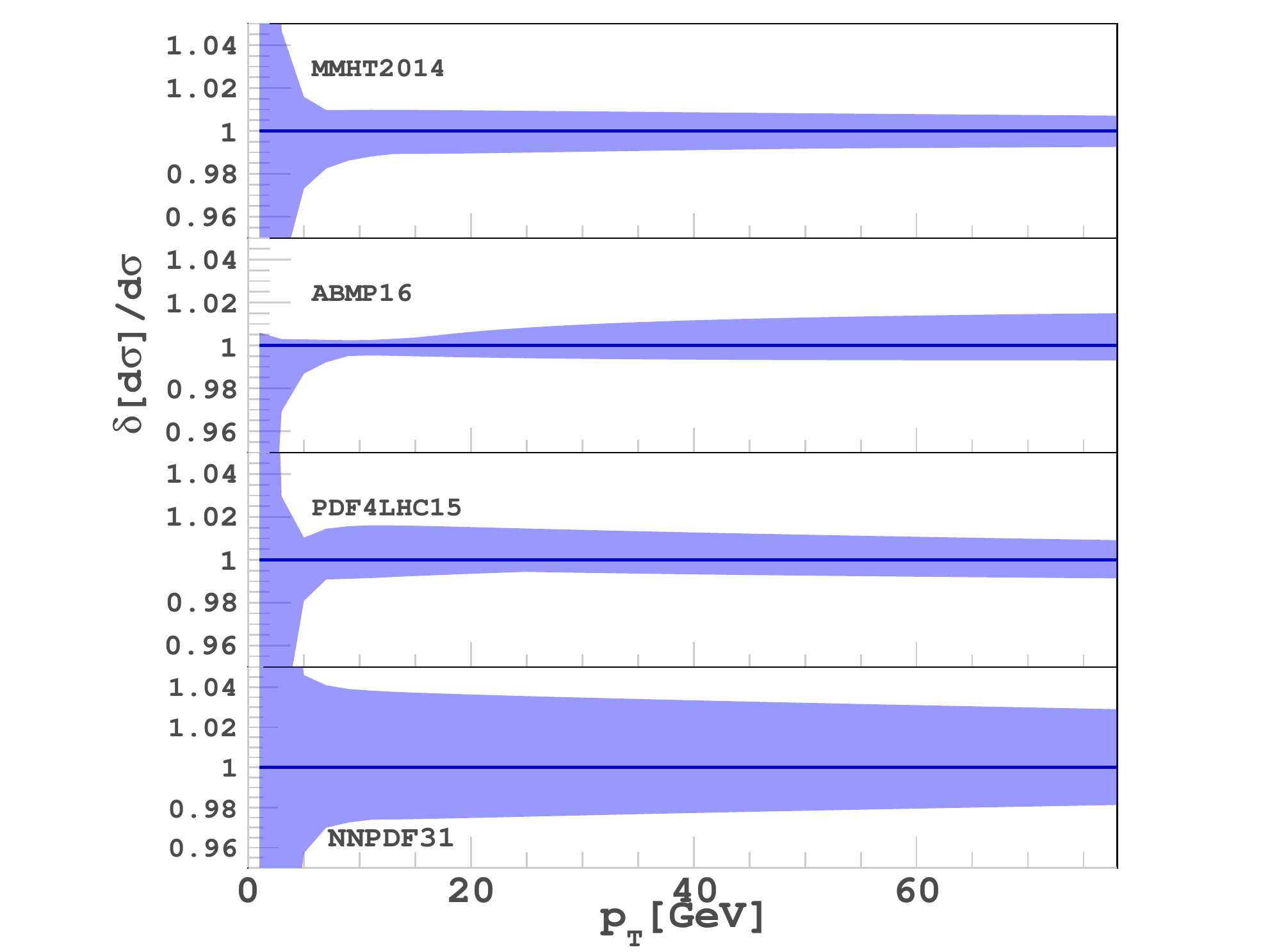}} &
     \mbox{\includegraphics[height=.25\textheight]{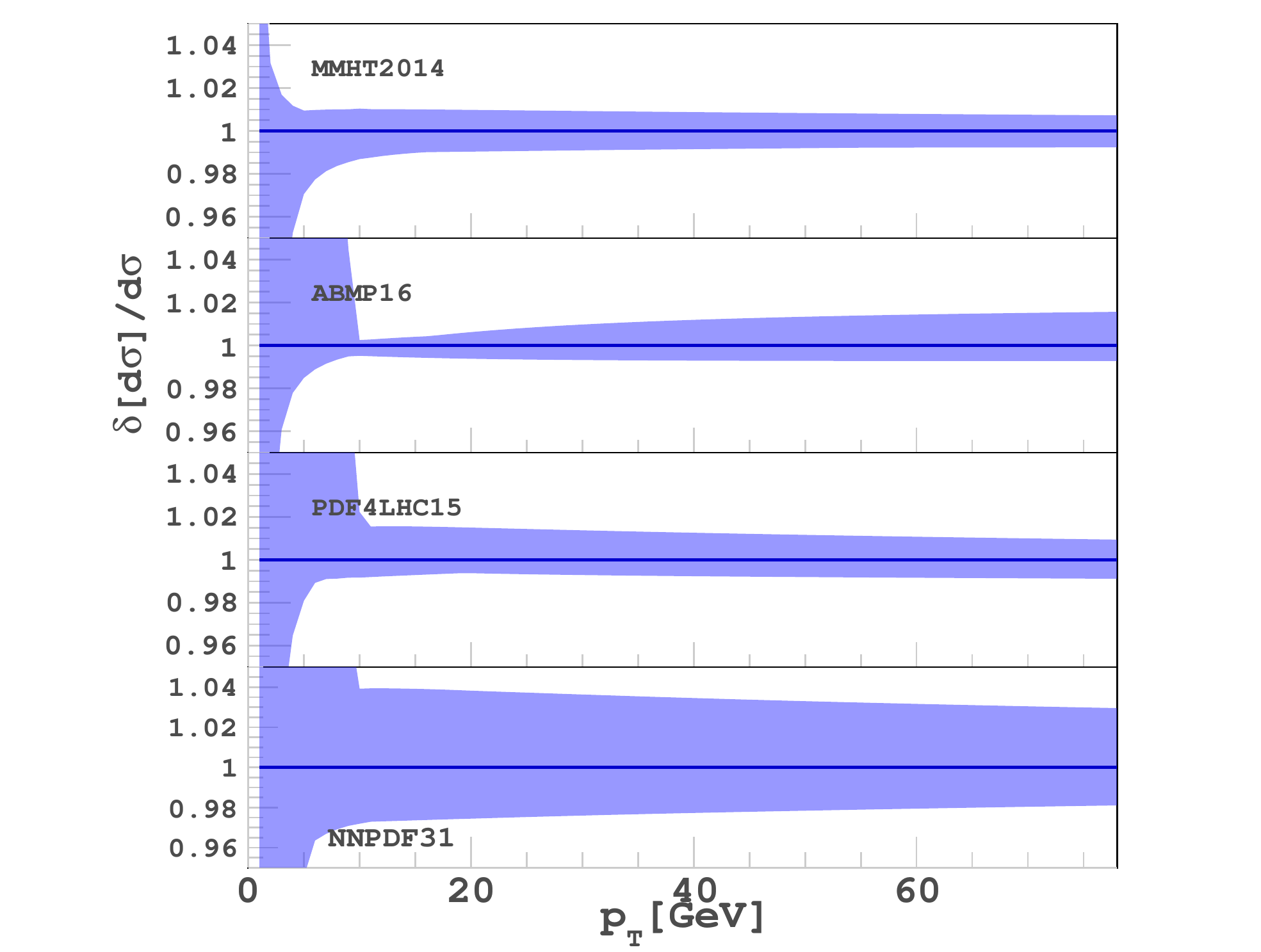}}
            \\
     \hspace{0.7cm} (a) &\qquad (b)
    \end{tabular}
    \parbox{.8\textwidth}{%
      \caption{\label{PDF}PDF variation at NNLO$_A$+NNLL  for (a) 14 TeV and (b) 13 TeV using various sets. The y-axis represents the ratio of extremum variation over the central PDF set.}}
\end{center}
\end{figure}

\paragraph{Uncertainty due to $Q$:}
in Fig.~\ref{Qvar1} and Fig.~\ref{Qvar2}  we also show the sensitivity of the resummed
 results to the choice of resummation scale $Q$, where we have varied $Q$ from $m_{A}$
 to $m_{A}/8$.
For each diagram, in the left panel we see the results are quite sensitive to the choice, where by
sensitivity we mean the range of variation of the maxima of distribution for different choices of Q.
 Not surprisingly, upon going 
to the next logarithmic accuracy (right panel) the sensitivity is significantly reduced around the peak
region and the results at moderate values of $\pt$ are almost insensitive to the choice. 
It is reassuring that in the right panel at moderate and large values of $\pt$ the resummed curve is coincident
with the fixed order curve, as desired.
We note that the position of the peak is unchanged in going to the next order. For $Q=m_A$ and centre-of-mass energy 14 TeV we see that the peak
value changes by 25\% in going from NLO+NLL to NNLO$_A$+NNLL. Similarly for $Q=m_A/2$ and centre-of-mass energy 13 TeV, the peak value changes by  11\% upon going from NLO+NLL  to the next level of accuracy.
\paragraph{Uncertainty due to $\mu_R$ and $\mu_F$:}
in Fig.~\ref{figure2}(a) and Fig.~\ref{figure2}(b)
we show the sensitivity of our results to the variation of $\mu_R$ and $\mu_F$. 
The bands in this figure have been obtained by varying of $\mu_R$ and $\mu_F$ independently in the range 
$[m_A/2, 2m_A]$, while excluding the regions where $\mu_R/\mu_F>2$ or $\mu_R/\mu_F<1/2$.
More specifically, for 14 TeV centre-of-mass energy we see
that at the peak, the variation is about $38\%$ for NLO+NLL which gets reduced to about $19\%$ upon going to the next level
of accuracy. Similarly for 13 TeV centre-of-mass energy we see
that at the peak the variation is about  $22\%$ for NLO+NLL  and about $15\%$  for NNLO$_A$+NNLL. We have also studied the individual variation of $\mu_R$ and $\mu_F$  for 
both the energies at the LHC in Fig.~\ref{figure3} and Fig.~\ref{figure4} respectively. {In Fig.~\ref{figure3} we keep $\mu_F=m_A$ and vary $\mu_R$ in the 
 range $[m_A/2, 2m_A]$. For 14 TeV centre-of-mass energy we find that at the peak, the variation for NLO+NLL is about $32\%$, which gets reduced to about $17\%$ at  NNLO$_A$+NNLL. Similarly for 13 TeV centre-of-mass energy we see
that at the peak the variation is about  $21\%$ for NLO+NLL  and about $13\%$  for NNLO$_A$+NNLL.
 In Fig.~\ref{figure4} we set $\mu_R=m_A$ and vary $\mu_F$ in the 
same range as above. For 14 TeV centre-of-mass energy we find that at the peak, the variation for NLO+NLL is about $4\%$, which gets reduced to about $2\%$ at  NNLO$_A$+NNLL. Similarly for 13 TeV centre-of-mass energy we see
that at the peak the variation is about  $4\%$ for NLO+NLL  and about $0.5\%$  for NNLO$_A$+NNLL.}
\paragraph{Combined uncertainty due to  $Q, \mu_R$ and $\mu_F$:} in Fig.~\ref{figure5}(a) and Fig.~\ref{figure5}(b)
we show the sensitivity of our results to the variation of Q, $\mu_R$ and $\mu_F$. The bands in this figure show independent variation of Q, $\mu_R$ and $\mu_F$ in the range $[m_A/2, 2m_A]$ with constraints $\mu_R/\mu_F\in[1/2,\,2]$, $Q/\mu_R\in[1/2,\,2]$ and $Q/\mu_F\in[1/2,\,2]$.
When we take into account all scale variations together we notice that both at 13 TeV and 14 TeV the variation at the peak is $38\%$ for NLO+NLL which gets reduced to $20\%$ upon going to the next level
of accuracy. It is to be noted that this amount of decrement is almost same as the case discussed in
Fig.~\ref{figure2}(a).
\paragraph{Uncertainty due to parton density sets:}
as there are several PDF groups in the literature, it is necessary to estimate the uncertainty resulting from the choice of PDFs within each set of a given PDF group.  Using PDFs from different PDF groups namely MMHT2014~\cite{Harland-Lang:2014zoa}, ABMP~\cite{Alekhin:2017kpj}, NNPDF3.1~\cite{Ball:2017nwa} and PDF4LHC~\cite{Butterworth:2015oua}  we have obtained the differential $p_T$ distributions along with the corresponding PDF uncertainty. In~Fig.~\ref{PDF}(a), 
we have demonstrated the uncertainty bands for various PDF sets as a function of $p_T$ at energies of 14 TeV.
In order to demonstrate the correlation of PDF uncertainty with the $p_T$ values  we have tabulated in~Table~\ref{table1}, the corresponding results for few benchmark values of $p_T$ along with percentage uncertainties. We have also performed the same exercise for 13 TeV centre of mass energy, as shown in Fig.~\ref{PDF}(b). We have tabulated the  results for few benchmark values of $p_T$ along with percentage uncertainties in~Table~\ref{table2}.

\begin{figure}[h!]
\begin{center}
    \begin{tabular}{cc}
     \mbox{\includegraphics[height=.23\textheight]{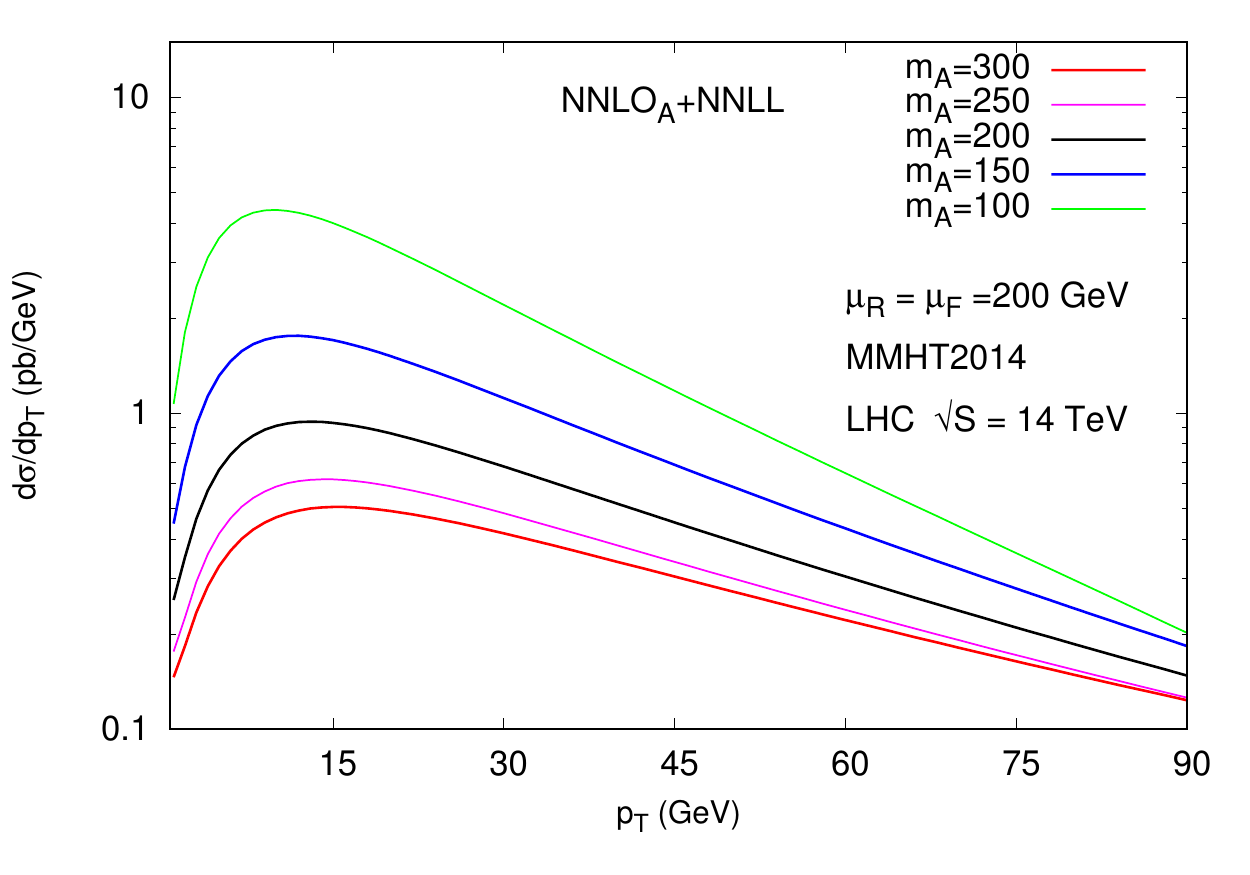}} &
     \mbox{\includegraphics[height=.23\textheight]{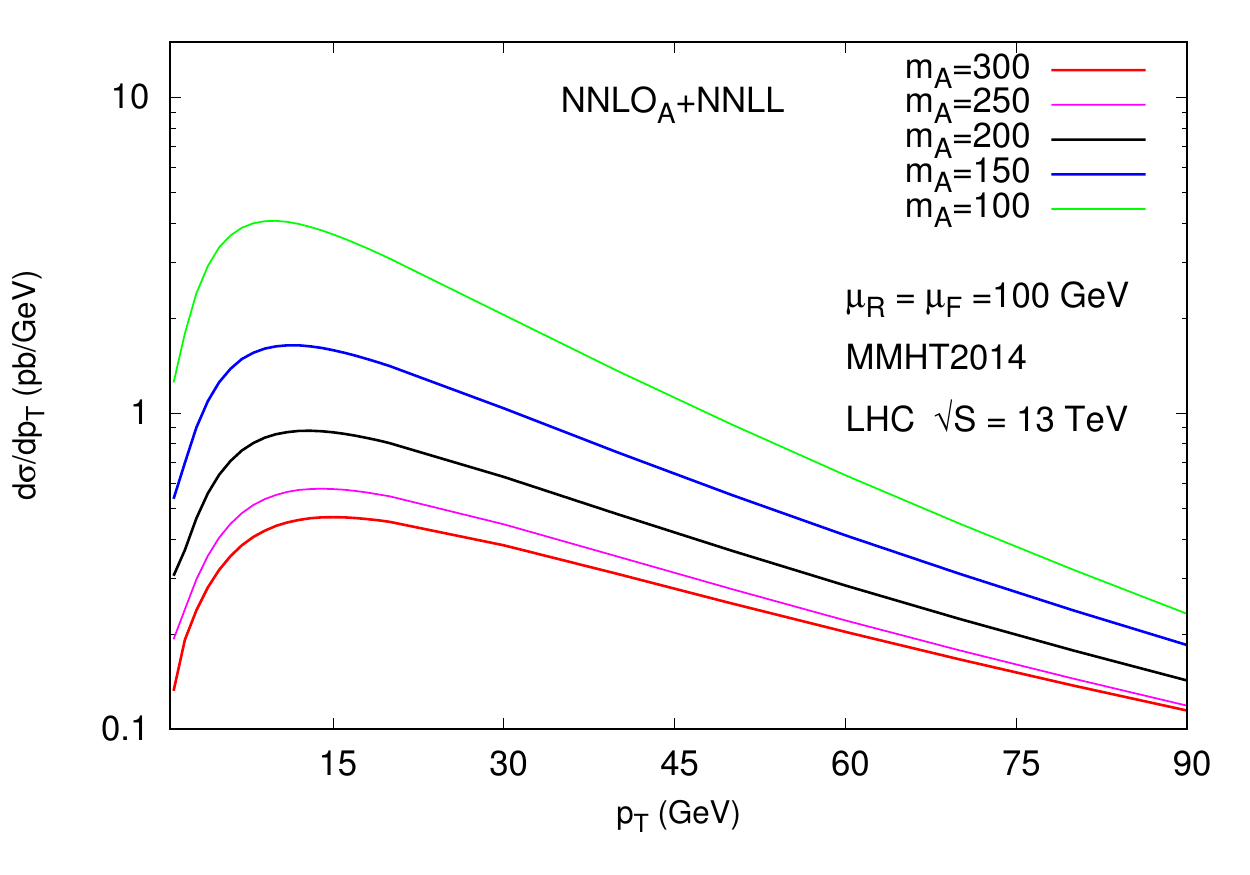}}
            \\
     \hspace{0.7cm} (a) &\qquad (b)
    \end{tabular}
    \parbox{.8\textwidth}{%
      \caption{\label{figure7}Pseudo-scalar Higgs mass variation
        at \nnlo{$_A$}\plus\nnll{}  for (a) 14 TeV and (b) 13 TeV}}
\end{center}
\end{figure}

\begin{table}[h!]
\centering
\renewcommand{\arraystretch}{1.7}
\begin{tabular}{|c|c|c|c|c|}
 \hline
  $q_T$ &MMHT  & ABMP  &NNPDF & PDF4LHC  \\
 \hline
\hline
  7.0 &  $0.802^{+0.97\%}_{-1.75\%}$  & $0.828^{+0.26\%}_{-0.78\%}$ & $0.821^{+4.09\%} _{-3.00\%}$&$0.804^{+1.45\%}_{-0.91\%}$ \\
  \hline
  13.0 &$0.941^{+0.98\%}_{-1.07\%}$ &$0.928^{+0.31\%}_{-0.49\%}$  &$0.960 ^{+3.77\%}_{-2.60\%}$&$0.943^{+1.60\%}_{-0.80\%}$  \\
  \hline
  19.0&$0.882^{+0.96\%}_{-1.05\%}$ &$0.847^{+0.58\%}_{-0.55\%} $ & $0.897^{+3.65\%}_{-2.53\%}$ &$0.884^{+1.54\%}_{-0.67\%} $\\
\hline
 25.0 &$0.772^{+0.94\%}_{-1.01\%}$ &$0.729^{+0.83\%}_{-0.60\%}$  &$0.783 ^{+3.55\%}_{-2.45\%}$&$0.774^{+1.46\%}_{-0.56\%}$  \\
  \hline
   31.0 &$0.660^{+0.91\%}_{-0.96\%}$ &$0.616^{+0.99\%}_{-0.63\%}$  &$0.669 ^{+3.46\%}_{-2.38\%}$&$0.662^{+1.38\%}_{-0.61\%}$  \\
  \hline
\end{tabular}
\caption{$q_T$ distributions at NNLO$_A$+NNLL using different PDF sets along with percentage uncertainties
for  $q_T= 7.0 , 13.0, 19.0, 25.0, 31.0$ for $\sqrt{s}=14$ TeV.}
\label{table1}
\end{table}
\begin{table}[h!]
\centering
\renewcommand{\arraystretch}{1.7}
\begin{tabular}{|c|c|c|c|c|}
 \hline
  $q_T$ &MMHT  & ABMP  &NNPDF & PDF4LHC  \\
 \hline
\hline
  7.0 &  $0.762^{+0.99\%}_{-1.87\%}$  & $0.783^{+18.52\%}_{-0.85\%}$ & $0.780^{+22.04\%} _{-3.30\%}$&$0.761^{+23.32\%}_{-0.89\%}$ \\
  \hline
  13.0 &$0.880^{+1.01\%}_{-1.11\%}$ &$0.864^{+0.34\%}_{-0.54\%}$  &$0.898 ^{+3.93\%}_{-2.67\%}$&$0.882^{+1.56\%}_{-0.75\%}$  \\
  \hline
  19.0&$0.820^{+0.98\%}_{-0.97\%}$ &$0.783^{+0.57\%}_{-0.61\%} $ & $0.834^{+3.84\%}_{-2.56\%}$ &$0.822^{+1.51\%}_{-0.62\%} $\\
\hline
 26.0 &$0.698^{+0.94\%}_{-0.92\%}$ &$0.654^{+0.85\%}_{-0.66\%}$  &$0.707 ^{+3.70\%}_{-2.45\%}$&$0.700^{+1.43\%}_{-0.68\%}$  \\
  \hline
   32.0 &$0.596^{+0.91\%}_{-0.89\%}$ &$0.552^{+0.99\%}_{-0.63\%}$  &$0.602 ^{+3.60\%}_{-2.37\%}$&$0.597^{+1.35\%}_{-0.72\%}$  \\
  \hline
\end{tabular}
\caption{$q_T$ distributions at NNLO$_A$+NNLL using different PDF sets along with percentage uncertainties
for  $q_T= 7.0 , 13.0, 19.0, 26.0, 32.0$ for $\sqrt{s}=13$ TeV.}
\label{table2}
\end{table}
\paragraph{Pseudo-scalar Higgs mass variation:}
in Fig.~\ref{figure7}(a) and Fig.~\ref{figure7}(b) 
we show how the distribution behaves as the mass of the final state is changed. We have kept the renormalisation 
and factorisaiton scales fixed at $200$ GeV for 14 TeV, at 100 GeV for 13 TeV LHC energies and varied $m_A$ from $100$ to $300$ GeV.   We see 
that the cross-section decreases with the increase in the mass of the final state.
\section{Conclusion} \label{conclusion}
In this study we obtained the resummed $\pt$ distribution for pseudo-scalar Higgs bosons at the LHC for both the centre-of-mass energy 14 TeV and 13 TeV
at next-to-next-to-leading logarithmic accuracy by matching the resummed curve with approximated fixed order next-to-next-to-leading order result.
We showed that we achieve a very significant reduction in sensitivity to the choices of resummation, renormalisation and factorisation 
scales that are artefact of perturbation theory. We also studied the uncertainty due to different choices of parton density sets.
These results provide us with precise estimate for the distribution especially in the region around $15$ GeV where the cross-section is 
large and the fixed order results are completely unreliable due to the breakdown of fixed order perturbation series. 

\section*{Acknowledgments}
 AT would like to thank G. Ferrera for very helpful discussions,
and  would also like to thank the Department of Physics of the University of Turin  for hospitality during the last stages of this work.
NA would like to thank B. L. Reddy for his support.
 GD would like to thank F. J. Tackmann for stimulating discussions and suggestions on the manuscript.
\bibliographystyle{JHEP}
\bibliography{qtresum}
\end{document}